\documentclass[10pt,final,comcos]{IEEEtran}

\usepackage[T1]{fontenc}

\usepackage{amsmath}
\usepackage{graphicx}

\usepackage{epstopdf}
\DeclareGraphicsExtensions{.pdf,.eps,.png,.jpg,.mps}

\usepackage{algorithm}

\usepackage{algorithmicx}

\usepackage{algpseudocode}

\usepackage{bbm}

\usepackage[tight,footnotesize]{subfigure}

\usepackage{amssymb}

\usepackage{epsfig}

\usepackage{stfloats}

\usepackage{endnotes}

\usepackage{multirow}

\usepackage{adjustbox}

\usepackage{subfigure}

\usepackage{enumerate}

\usepackage{enumitem}

\DeclareFontFamily{OT1}{pzc}{}
\DeclareFontShape{OT1}{pzc}{m}{it}{<-> s * [1.10] pzcmi7t}{}
\DeclareMathAlphabet{\mathpzc}{OT1}{pzc}{m}{it}

\makeatletter
\def\user@resume{resume}
\def\user@intermezzo{intermezzo}
\newcounter{previousequation}
\newcounter{lastsubequation}
\newcounter{savedparentequation}
\renewenvironment{subequations}[1][]{%
      \def\user@decides{#1}%
      \setcounter{previousequation}{\value{equation}}%
      \ifx\user@decides\user@resume
           \setcounter{equation}{\value{savedparentequation}}%
      \else
      \ifx\user@decides\user@intermezzo
           \refstepcounter{equation}%
      \else
           \setcounter{lastsubequation}{0}%
           \refstepcounter{equation}%
      \fi\fi
      \protected@edef\theHparentequation{%
          \@ifundefined {theHequation}\theequation \theHequation}%
      \protected@edef\theparentequation{\theequation}%
      \setcounter{parentequation}{\value{equation}}%
      \ifx\user@decides\user@resume
           \setcounter{equation}{\value{lastsubequation}}%
         \else
           \setcounter{equation}{0}%
      \fi
      \def\theequation  {\theparentequation  \alph{equation}}%
      \def\theHequation {\theHparentequation \alph{equation}}%
      \ignorespaces
}{%
  \ifx\user@decides\user@resume
       \setcounter{lastsubequation}{\value{equation}}%
       \setcounter{equation}{\value{previousequation}}%
  \else
  \ifx\user@decides\user@intermezzo
       \setcounter{equation}{\value{parentequation}}%
  \else
       \setcounter{lastsubequation}{\value{equation}}%
       \setcounter{savedparentequation}{\value{parentequation}}%
       \setcounter{equation}{\value{parentequation}}%
  \fi\fi
  \ignorespacesafterend
}
\makeatother

\makeatletter
\newcommand{\subalign}[1]{%
  \vcenter{%
    \Let@ \restore@math@cr \default@tag
    \baselineskip\fontdimen10 \scriptfont\tw@
    \advance\baselineskip\fontdimen12 \scriptfont\tw@
    \lineskip\thr@@\fontdimen8 \scriptfont\thr@@
    \lineskiplimit\lineskip
    \ialign{\hfil$\m@th\scriptstyle##$&$\m@th\scriptstyle{}##$\crcr
      #1\crcr
    }%
  }
}
\makeatother

\newtheorem{lemma}{Lemma}

\newtheorem{theorem}{Theorem}

\newtheorem{definition}{Definition}

\begin{document}


\title{
Dynamic Power Splitting for SWIPT with Nonlinear Energy Harvesting in Ergodic Fading Channel
}

\author{
Jae-Mo~Kang, Chang-Jae Chun, Il-Min~Kim,~\IEEEmembership{Senior~Member,~IEEE}, and Dong~In~Kim,~\IEEEmembership{Senior~Member,~IEEE} 
\thanks{J.-M. Kang, C.-J. Chun, and I.-M. Kim are with the Department of Electrical and Computer Engineering, Queen's University, Kingston, ON K7L 3N6, Canada (e-mail: jaemo.kang@queensu.ca; changjae.chun@queensu.ca; ilmin.kim@queensu.ca).}
\thanks{D. I. Kim is with the School of Information and Communication Engineering, Sungkyunkwan University (SKKU), Suwon, 16419, South Korea (e-mail: dikim@skku.ac.kr).}
}

\maketitle

\begin{abstract}
We study the dynamic power splitting for simultaneous wireless information and power transfer (SWIPT) in the ergodic fading channel.
Considering the nonlinearity of practical energy harvesting circuits, we adopt the realistic nonlinear energy harvesting (EH) model rather than the idealistic linear EH model.
To characterize the optimal rate-energy (R-E) tradeoff, we consider the problem of maximizing the R-E region, which is nonconvex.
We solve this challenging problem for two different cases of the channel state information (CSI):
(i) when the CSI is known only at the receiver (the CSIR case) and (ii) when the CSI is known at both the transmitter and the receiver (the CSI case).
For these two cases, we develop the corresponding optimal dynamic power splitting schemes. To address the complexity issue, we also propose the suboptimal schemes with low complexities.
Comparing the proposed schemes to the existing schemes, we provide various useful and interesting insights into the dynamic power splitting for the nonlinear EH.
Furthermore, we extend the analysis to the scenarios of the partial CSI at the transmitter and the harvested energy maximization.
Numerical results demonstrate that the proposed schemes significantly outperform the existing schemes
and the proposed suboptimal scheme works very close to the optimal scheme at a much lower complexity.
\end{abstract}


\IEEEpeerreviewmaketitle

\section{Introduction}

As a promising technology for future energy-constrained networks,
simultaneous wireless information and power transfer (SWIPT) using radio frequency (RF) signals has recently drawn an upsurge of research interest in the literature \cite{Lu15}.
A practical limitation for the SWIPT is that with the current circuit technology, it is difficult to carry out both information decoding (ID) and energy harvesting (EH) at the same time from the same received RF signal \cite{Zhang13}, \cite{Zhou13}.
Thus, in the practical SWIPT system, there exists a tradeoff between the amount of information transfer and the amount of energy transfer, which is called the rate-energy (R-E) tradeoff \cite{Zhang13}, \cite{Zhou13}.
To characterize and understand the fundamental performance of the SWIPT system, analyzing the R-E tradeoff is essential and crucial.
In the literature, the R-E tradeoff for the SWIPT has been studied under different fading environments, e.g.,
in the additive white Gaussian noise (AWGN) channel without fading \cite{Zhang13}--\cite{Xiong} and in the fading channel \cite{Kang_2}--\cite{Liu_dps}.
Unlike the AWGN channel, in the fading channel, the received RF power variation due to the fading fluctuation can be effectively and opportunistically utilized for the SWIPT.
There are two main different schemes for this purposes: the mode switching and the dynamic power splitting.
In almost all the works for the SWIPT in the fading channel including \cite{Kang_2}--\cite{Lu}, the mode switching was studied.
On the other hand, the study on dynamic power splitting in the fading channel is rare despite its generality and superior performance: 
to the best of our knowledge, the issue was studied \textit{only} in \cite{Liu_dps}.

In \cite{Liu_dps}, the optimal dynamic power splitting scheme was developed in the sense of the R-E tradeoff.
However, in \cite{Liu_dps}, it was assumed that the amount of energy harvested by the EH circuit is linearly proportional to the received RF power, namely, the linear EH model.
This idealistic (i.e., linearity) assumption is valid only when the energy conversion efficiency of the EH circuitry is the same (i.e., constant) for the infinitely wide received RF power level.
However, as studied in the very recent literature \cite{Kang}--\cite{Kang_3}, \cite{Clerckx_0}--\cite{Boshkovska17_2},
the linear EH model is overly idealistic and unrealistic because the linearity assumption does not hold in practice.
Specifically, as validated in many experimental results \cite{Valenta14}--\cite{Guo12}, the energy conversion efficiency of the practical EH circuit becomes different (not constant) depending on the received RF power level.
Also, as analyzed in \cite{Clerckx_0}--\cite{Clerckx_2}, the energy conversion efficiency of the actual EH circuit (or rectifier) is a nonlinear function of the received RF signal (i.e., power and shape)
due to various causes of nonlinearity, e.g., nonlinearity of the diode.
In practice, therefore, the amount of harvested energy is clearly a nonlinear function of the received power.
Unfortunately, the linear EH model cannot accurately model the nonlinear behavior of the practical EH circuit \cite{Kang}--\cite{Kang_3}, \cite{Clerckx_0}--\cite{Boshkovska17_2},
which may incur severe mismatch or inaccuracy in the practical system.
To overcome the limitations of the linear model, one should address the practical issue of nonlinear EH.


In \cite{Kang_2}, \cite{Kang_3}, the mode switching scheme was studied for the realistic scenario of nonlinear EH.
In these works, an important and crucial conclusion was made: the mode switching scheme developed for the linear EH is no longer optimal for the nonlinear EH,
because the linear and nonlinear EH models are practically and mathematically different.
From this result, one can expect that the existing dynamic power splitting scheme developed in \cite{Liu_dps} for the linear EH might not work well or might lead to misleading/wrong conclusion for the nonlinear EH.
In addition, none of the existing mode switching schemes developed in \cite{Kang_2}, \cite{Kang_3} for the nonlinear EH are truly optimal in the sense of achieving the ultimate R-E tradeoff performance of the SWIPT system. This is because the mode switching can be considered as a \textit{special} and \textit{simplified} version of the dynamic power splitting \cite{Liu_dps},
meaning that the dynamic power splitting generally yields better performance than the mode switching.
To achieve the theoretically best performance of the SWIPT system in the fading channel, the corresponding optimal dynamic power splitting scheme must be studied.
To the best of our knowledge, for the SWIPT system with nonlinear EH in the ergodic fading channel, the optimal dynamic power splitting scheme has not been studied in the literature.
This motivated our work.

In this paper, we study the dynamic power splitting for the SWIPT system with nonlinear EH in the ergodic fading channel.
For analysis, we adopt a realistic nonlinear EH model developed in \cite{Boshkovska15}--\cite{Boshkovska17_2},
which was shown to accurately match the experimental results \cite{X_Xu}.
Using this nonlinear model, to characterize the optimal R-E tradeoff,
we formulate the optimal dynamic power splitting problem to maximize the average achievable rate
under the constraints on the average harvested energy and the average transmit power,
which is a nonconvex problem. We solve this challenging problem for three different cases of the channel state information (CSI):
(i) when the instantaneous CSI is known only at the receiver (the CSIR case), 
(ii) when the partial and instantaneous CSI is known at the transmitter and the receiver, respectively (the case of partial CSI at the transmitter),
and (iii) when the instantaneous CSI is known at both the transmitter and the receiver (the CSI case).
The main contributions of this paper are as follows:
\begin{itemize}
\item In the case of CSIR, we develop the optimal power splitting scheme for the SWIPT system with nonlinear EH. Also, to address the complexity issue of the optimal scheme, we propose a suboptimal scheme with low complexity,
    which is shown to be asymptotically optimal in the low signal-to-noise ratio (SNR) regime.
    
\item Furthermore, we extend the analysis for the CSIR case to the case of partial CSI at the transmitter.

\item As a different case from the CSIR case, we further study the CSI case. In this case, we develop the optimal and suboptimal power splitting schemes.
We also establish the asymptotic optimality of the proposed suboptimal scheme in the low SNR regime.
Furthermore, we extend our analysis for the rate maximization to the case of harvested energy maximization.

\item Through comparisons between the proposed and existing schemes, we provide various useful and interesting insights into the dynamic power splitting for the nonlinear EH.
\end{itemize}

This paper is organized as follows.
In Section II, the system model is described and the problem is formulated.
In Section III, we develop the optimal and suboptimal dynamic power splitting schemes for the CSIR case
and we extend the analysis to the case of partial CSI at the transmitter.
In Section IV, the analysis for the CSI case is presented.
Section V presents the numerical results and Section VI concludes the paper.

\section{System Model and Problem Formulation}

We consider a point-to-point SWIPT system with a transmitter and a receiver, each of which is equipped with a single antenna.
It is assumed that the channel between the transmitter and receiver remains constant during one coherent fading block of duration $T$,
but it varies from one block to another block independently \cite{Tse}.
We assume the ergodic fading channel: a codeword spans over many fading blocks, i.e., $N \rightarrow \infty$, where $N$ denotes the number of fading blocks.

The dynamic power splitting scheme is adopted at the receiver.\footnote{
Different from the original dynamic power splitting scheme developed in \cite{Zhou13} where the power splitting ratio was assumed to vary at the symbol level,
in this paper, we assume that the power splitting ratio varies at the coherent block level.
Thanks to much less frequent changes of the power splitting ratio,
the proposed dynamic power splitting scheme is much simpler than that of \cite{Zhou13}, and thus, it can be implemented much easily and will be highly likely suitable for real applications.
}
At a particular fading state $\nu$, the received power is dynamically split with a power splitting ratio $0 \leq \rho_{\nu} \leq 1$.
Specifically, the $ \rho_{\nu} $ portion of the received power is used for EH
and the remaining $(1 -  \rho_{\nu} ) $ portion of the received power is used for ID.
Let $P_{\nu}$ denote the transmit power at the fading state $\nu$.
Then the average achievable rate over the fading blocks is given by $\mathbb{E} \left[ R_{\nu} (  P_{\nu} , \rho_{\nu}  )  \right]$ \cite{Tse},
where
$\mathbb{E} [ \cdot ]$ denotes the expectation operation taken over the fading process
and
\begin{align}
\label{R}
R_{\nu} ( P_{\nu}, \rho_{\nu}   )  =  \log_{2} \left(  1  +  \frac{ (1 - \rho_{\nu}) h_{\nu} P_{\nu} }{ \sigma^{2} }  \right).
\end{align}
Also, $ h_{\nu} $ denotes the channel power gain at the fading state $\nu$, and $\sigma^{2}$ the variance of the additive noise.
In the next subsection, the amount of harvested energy will be discussed in detail.

\subsection{Linear and Nonlinear Energy Harvesting}

In the existing works including \cite{Lee}--\cite{Lu}, \cite{Liu_dps}, the linear EH model was adopted. In the linear EH model, at the fading state $\nu$,
the amount of harvested energy is modeled as
\begin{align}
\label{Q_L}
Q_{\nu}^{\texttt{L}} ( P_{\nu} , \rho_{\nu} )  =  \zeta \rho_{\nu} h_{\nu} P_{\nu} T.
\end{align}
In (\ref{Q_L}), $0 < \zeta \leq 1$ denotes the energy conversion efficiency, which is assumed to be a constant and be independent of the received RF power $ h_{\nu} P_{\nu} $.
However, the energy conversion efficiency of the actual EH circuit is different (not constant) over the different received RF power levels,
as studied in \cite{Clerckx_0}--\cite{Clerckx_2} and validated in the experimental results \cite{Valenta14}--\cite{Guo12}.
In practice, the amount of harvested energy of the actual EH circuit increases \textit{nonlinearly} with the received RF power.
Specifically, in the low RF power level, the energy conversion efficiency is very small (close to zero) due to the turn-on voltage of the diode; in the middle RF power level, the efficiency is large (about 0.7 at the RF frequency of 915 MHz \cite{Valenta14}) because the diode works in the linear region; and in the high RF power level, the efficiency is again very small (close to zero) due to the reverse breakdown of the diode.
Unfortunately, the simplistic linear model of (\ref{Q_L}) cannot accurately model the nonlinearity of the practical EH circuits,
and thus, it is never realistic in practical systems.

In order to address the critical limits of the linear EH model
and to accurately model the nonlinear behavior of the actual EH circuit,
several realistic nonlinear EH models were developed and studied in the recent literature \cite{Clerckx_0}--\cite{Boshkovska17_2}.
In \cite{Clerckx_0}--\cite{Clerckx_2}, the nonlinearity of the rectifier was modeled based on the nonlinearity of the diode characteristic equation.
On the other hand, in \cite{X_Xu}--\cite{Boshkovska17_2}, the nonlinearity of the energy conversion efficiency was modeled based on nonlinear functions.
Among the various nonlinear models, the nonlinear model developed in \cite{Boshkovska15}--\cite{Boshkovska17_2} was shown to accurately match the experimental results, e.g., see \cite[Figs. 2, 3, and Table I]{X_Xu}, \cite[Fig. 2]{Boshkovska15}. In this paper, for accuracy, validity, and practicality of the analysis with useful insights, we adopt the realistic nonlinear EH model of \cite{Boshkovska15}--\cite{Boshkovska17_2}.
Note that the analysis presented in this paper can be extended to the other nonlinear models considered in \cite{Zeng}--\cite{X_Xu}.
In the adopted nonlinear model, the amount of the harvested energy is modeled using the logistic (or sigmoid) function, i.e., S-shaped curve, as follows \cite{Boshkovska15}--\cite{Boshkovska17_2}:
\begin{align}
\label{Q_NL}
Q_{\nu}^{\texttt{NL}} (  P_{\nu} , \rho_{\nu}  )  =   \frac{  P_{s} T \left[ \Psi_{\nu} (  P_{\nu} , \rho_{\nu} )  -  \Omega \right]  }{ 1 - \Omega }
\end{align}
where $ \Omega = \frac{ 1 }{ 1 + e^{  a b } }  $ is a constant to ensure the zero-input zero-output response
and $\Psi_{\nu} ( P_{\nu}, \rho_{\nu}  )$ is the logistic function given by
\begin{align}
\label{Psi}
\Psi_{\nu} ( P_{\nu}, \rho_{\nu}  )  =  \frac{ 1 }{ 1 + e^{ - a (  \rho_{\nu}  h_{\nu} P_{\nu} - b ) } }.
\end{align}
In (\ref{Q_NL}), $P_{s}$ denotes the maximum amount of harvested power when the EH circuit is saturated.
Also, $a$ and $b$ are positive constants related to the circuit specification:
$a$ represents the nonlinear charging rate with respect to the input power and $b$ is related to the turn-on threshold.
\subsection{Problem Formulation}

In this paper, using the realistic nonlinear model $Q_{\nu}^{\texttt{NL}} (\cdot)$, we aim to develop the optimal dynamic power splitting scheme in the sense of the R-E tradeoff.
We consider two different cases of the CSI availability: (i) the CSIR case and (ii) the CSI case.
In the following, the optimization problems for these two cases are formulated.

\subsubsection{CSIR Case} In this case, the CSI is known at the receiver, but unknown at the transmitter.
Over the duration of a codeword, the receiver dynamically splits the received power using the power splitting ratios $0 \leq \rho_{\nu}  \leq 1$, $\forall \nu$.
On the other hand, the transmit power is fixed, i.e., $P_{\nu} = P$, $\forall \nu$.
In the case of CSIR, the R-E region is given by
\begin{align}
\label{R_E_region}
\mathcal{C}_{\rm CSIR}^{\texttt{NL}}  = \bigcup_{ 0 \leq \rho_{\nu} \leq 1,  \forall \nu  }  \Big\{ (R,Q): & ~ Q \leq \mathbb{E} \left[  Q_{\nu}^{\texttt{NL}} (  P , \rho_{\nu}  )  \right] , \nonumber\\
 & ~ R \leq \mathbb{E} \left[  R_{\nu} ( P , \rho_{\nu}  )  \right]  \Big\},
\end{align}
which contains all pairs of the average achievable rate and the average harvested energy.
To achieve the optimal R-E tradeoff with CSIR, we have to maximize the R-E region $\mathcal{C}_{\rm CSIR}^{\texttt{NL}}$ of (\ref{R_E_region}) by optimizing the power splitting ratios $\{ \rho_{\nu} \}$,
which can be formulated as follows:
\begin{subequations}
\label{P1}
\begin{align}
\label{P1_obj}
{\rm (P1):}  \quad  \underset{ 0 \leq \rho_{\nu} \leq 1,  \forall \nu }{\texttt{max}} & \quad \mathbb{E} \left[  R_{\nu} ( P , \rho_{\nu}  )  \right]   \\
\label{P1_const_1}
\texttt{s.t.}  &  \quad  \mathbb{E} \left[  Q_{\nu}^{\texttt{NL}} (  P , \rho_{\nu}  )  \right] \geq Q.
\end{align}
\end{subequations}
In (P1), $ 0 \leq Q \leq Q_{\max}^{\rm CSIR} $ denotes a threshold for the average harvested energy,
where $Q_{\max}^{\rm CSIR}  = \mathbb{E} \left[  Q_{\nu}^{\texttt{NL}} (  P , 1  )  \right]$ is the maximum amount of harvested energy achieved with $\rho_{\nu} = 1$, $\forall \nu$.

\subsubsection{CSI Case} Different from the CSIR case, we also study the CSI case assuming that the CSI feedback from the receiver to the transmitter is available.
In this case, the CSI is known at both the transmitter and the receiver.
Over the duration of a codeword, the receiver dynamically splits the received power using the power splitting ratios $0 \leq \rho_{\nu}  \leq 1$, $\forall \nu$.
At the same time, the transmitter dynamically adapts the transmit power $P_{\nu} $, $\forall \nu$, under the long-term power constraint $\mathbb{E} \left[ P_{\nu} \right] \leq P_{\rm avg}$ and the short-term power constraint $P_{\nu} \leq P_{\rm max}$, $\forall \nu$, where $P_{\rm avg}$ and $ P_{\max} $ denote the long-term and short-term power thresholds, respectively.
Thus, the R-E region is given by
\begin{align}
\label{R_E_region_2}
\mathcal{C}_{\rm CSI}^{\texttt{NL}}  = \bigcup_{ \subalign{ & 0 \leq \rho_{\nu} \leq 1,  \forall \nu, \\ & 0 \leq P_{\nu} \leq P_{\max} , \forall \nu, \\ & \mathbb{E} \left[  P_{\nu}  \right]  \leq P_{\rm avg}  }  }  \Big\{ (R,Q): & ~ Q \leq \mathbb{E} \left[  Q_{\nu}^{\texttt{NL}} (  P_{\nu} , \rho_{\nu}  )  \right] , \nonumber\\
& ~ R \leq \mathbb{E} \left[  R_{\nu} ( P_{\nu} , \rho_{\nu}  )  \right]  \Big\}.
\end{align}
To achieve the optimal R-E tradeoff with CSI, we have to maximize the R-E region $\mathcal{C}_{\rm CSI}^{\texttt{NL}}$ of (\ref{R_E_region_2}) by jointly optimizing the transmit power $\{ P_{\nu} \}$ and the power splitting ratios $\{ \rho_{\nu} \}$,
which can be formulated as follows:\footnote{
With CSI at the transmitter, it is also practically important to consider the instantaneous R-E tradeoff performance optimization.
This optimization can be considered as a special case of our long-term R-E tradeoff optimization of (P2) when only a particular fading state (or a single fading block) is considered,
and thus, its solution can be obtained from the results in Theorem \ref{thm_3}.
}
\begin{subequations}
\label{P2}
\begin{align}
\label{P2_obj}
{\rm (P2):}  \quad  \underset{ \subalign{ & 0 \leq \rho_{\nu} \leq 1,  \forall \nu, \\ & 0 \leq P_{\nu} \leq P_{\max} , \forall \nu  } }{\texttt{max}} & \quad \mathbb{E} \left[  R_{\nu} ( P_{\nu} , \rho_{\nu}  )  \right]   \\
\label{P2_const_1}
\texttt{s.t.}  &  \quad  \mathbb{E} \left[  Q_{\nu}^{\texttt{NL}} (  P_{\nu} , \rho_{\nu}  )  \right] \geq Q,  \\
\label{P2_const_2}
&  \quad  \mathbb{E} \left[  P_{\nu}  \right]  \leq P_{\rm avg}.
\end{align}
\end{subequations}
In (P2), $ 0 \leq Q \leq Q_{\max}^{\rm CSI} $ is the threshold for the average harvested energy,
where $ Q_{\max}^{\rm CSI} =  \underset{  0 \leq P_{\nu} \leq P_{\max} , \forall \nu   }{\texttt{max}}  \mathbb{E} \left[  Q_{\nu}^{\texttt{NL}} (  P_{\nu} , 1 )  \right] $, which can be determined based on the result of \cite[Theorem 1]{Kang_3}.
In (P2), to avoid any trivial solution for the power allocation, we assume that $ P_{\max} > P_{\rm avg}$.

Note that the problems (P1) and (P2) are nonconvex because the objective functions and the constraints are nonconvex.
Also, the objective and constraint functions of (P1) and (P2) involve the expectations over the fading process, of which closed-form expressions are very difficult to obtain.
Thus, it is generally very challenging to tackle the problems (P1) and (P2).
One might try to use the exhaustive searching to find the solution. However, this approach appears to be practically infeasible due to extremely high computational complexity that grows exponentially with the (large) number $N$ of fading blocks.
Also, such approach does not provide any insight.
To overcome the challenges, in the next two sections, we present the optimal solutions to (P1) and (P2) efficiently
by exploiting the time-sharing properties of (P1) and (P2).\footnote{
In this paper, to solve (P1) and (P2), we use the same analytical approach (i.e., exploiting the time-sharing property) as in \cite{Kang_2}, \cite{Kang_3}, \cite{Liu_dps}.
However, even when the time-sharing property is exploited, our problems (P1) and (P2) are much more difficult to solve than the problems studied in \cite{Kang_2}, \cite{Kang_3}, \cite{Liu_dps}
because the Lagrangian dual function of (P1) is nonconvex in $\{  \rho_{\nu} \}$,
and that of (P2) are jointly nonconvex in $\{  P_{\nu} \}$ and $\{  \rho_{\nu} \}$.
}

\section{Dynamic Power Splitting for Nonlinear EH with CSIR}
\label{sec_P_EH}

In this section, we first derive the optimal solution to (P1).
Then we propose a suboptimal solution to (P1) with low complexity.
Finally, we compare the proposed scheme to the existing schemes.


\subsection{Optimal Solution to (P1)}

In order to solve the problem (P1) optimally and efficiently,
we exploit the time-sharing condition proposed in \cite{Yu}.
To this end, in the following, we first define the time-sharing condition for (P1).
\begin{definition}[\textit{\cite[Definition 1]{Yu}}]
Let $\{  \rho_{ x, \nu}^{*} \} $ and $\{  \rho_{y, \nu}^{*} \}$ denote the optimal solutions to the problem (P1) with $ Q =  Q_{x}  $ and $Q = Q_{y}$, respectively. Then the problem (P1) is said to satisfy the \textit{time-sharing} condition (or time-sharing property) if the following condition holds: there always exists a feasible point $\{ \rho_{z, \nu} \}$
satisfying $\mathbb{E} [  Q_{\nu}^{\texttt{NL}} (  P, \rho_{z, \nu}   )  ] \geq \theta Q_{x} + (1 - \theta) Q_{y}$
and $ \mathbb{E} [ R_{\nu} ( P , \rho_{z, \nu}) ] \geq  \theta \mathbb{E} [ R_{\nu} ( P , \rho_{x, \nu}^{*}) ] + (1 - \theta) \mathbb{E} [ R_{\nu} ( P , \rho_{y, \nu}^{*}) ] $ for any $Q_{x}$, $Q_{y}$, and $0 \leq \theta \leq 1$.
\null \hfill $\blacksquare$
\end{definition}

A useful fact is that if an optimization problem satisfies the time-sharing condition,
then the strong duality always holds, i.e., the duality gap is always zero, regardless of the convexity of the problem \cite[Theorem 1]{Yu}.
In the following, we establish the time-sharing property of (P1).
\begin{lemma}
\label{lem_1}
In the problem (P1), the time-sharing condition is satisfied.
\end{lemma}
\begin{IEEEproof}
See Appendix \ref{proof_lems_1_2}.
\end{IEEEproof}

\begin{figure*}[t]
\begin{align}
\label{rho_NL}
\rho_{\nu}^{\texttt{NL}}  = \begin{cases}  0 ,   &   \rm{for~Case~1}  \\
    \begin{cases}  \rho_{{\rm o}, \nu}^{\texttt{NL}} ,  &  {\rm if} ~  R_{\nu} ( P, \rho_{{\rm o}, \nu}^{\texttt{NL}} )  + \lambda^{\texttt{NL}} Q_{\nu}^{\texttt{NL}} ( P, \rho_{{\rm o}, \nu}^{\texttt{NL}} ) >  R_{\nu} ( P , 0 )   \\ 0, & {\rm otherwise}  \end{cases},  & \rm{for~Case~2}  \\
    \begin{cases}  1 ,  &  {\rm if}  ~   \lambda^{\texttt{NL}} Q_{\nu}^{\texttt{NL}} ( P , 1 ) >  R_{\nu} ( P , 0 )  \\ 0, & {\rm otherwise}  \end{cases},  & \rm{for~Case~3} \\
    \rho_{{\rm o}, \nu}^{\texttt{NL}}   ,   &   \rm{for~Case~4}  \\
    1  ,   &   \rm{for~Case~5}
\end{cases}.  \tag{9}
\end{align}
\hrulefill
\begin{align}
\label{rho_NL_cases}
& \gamma ( h_{\nu} )  \geq \tfrac{\lambda^{\texttt{NL}}}{4}   && \quad {\rm for ~ Case ~ 1} \nonumber \\
& \lambda^{\texttt{NL}} f ( 0 ) \leq g ( h_{\nu} ) , ~   \lambda^{\texttt{NL}} f ( h_{\nu} )  <  g ( 0 ) ,  ~ {\rm and} ~  \gamma ( h_{\nu} ) < \tfrac{\lambda^{\texttt{NL}}}{4}  && \quad {\rm for ~ Case ~ 2} \nonumber\\
& \lambda^{\texttt{NL}} f ( 0 ) \leq g ( h_{\nu} ) ~ {\rm and} ~   \lambda^{\texttt{NL}} f ( h_{\nu} )  \geq  g ( 0 )   && \quad {\rm for ~ Case ~ 3} \nonumber\\
& \lambda^{\texttt{NL}} f ( 0 ) > g ( h_{\nu} ) , ~   \lambda^{\texttt{NL}} f ( h_{\nu} )  <  g ( 0 ) ,  ~ {\rm and} ~  \gamma ( h_{\nu} ) < \tfrac{\lambda^{\texttt{NL}}}{4}  && \quad {\rm for ~ Case ~ 4} \nonumber\\
& \lambda^{\texttt{NL}} f ( 0 ) > g ( h_{\nu} ) ~ {\rm and} ~ \lambda^{\texttt{NL}} f ( h_{\nu} ) \geq  g ( 0 )  && \quad {\rm for ~ Case ~ 5}. \tag{11}
\end{align}
\hrulefill
\begin{align}
\label{rho_NL_so}
\rho_{\nu}^{\texttt{NL}'}  = \begin{cases}  0 ,   &   \rm{for~Case~1'}  \\
    \begin{cases}  \rho_{{\rm so}, \nu}^{\texttt{NL}} ,  &  {\rm if} ~ (1 - \rho_{{\rm so}, \nu}^{\texttt{NL}}) R_{\nu} ( P, 0 )  + \lambda^{\texttt{NL}} Q_{\nu}^{\texttt{NL}} ( P, \rho_{{\rm so}, \nu}^{\texttt{NL}} ) > R_{\nu} ( P, 0 )   \\ 0, & {\rm otherwise}  \end{cases},  & \rm{for~Case~2'}  \\
    \begin{cases}  1 ,  &  {\rm if}  ~ \lambda^{\texttt{NL}} Q_{\nu}^{\texttt{NL}} ( P, 1 ) > R_{\nu} ( P, 0 )  \\ 0, & {\rm otherwise}  \end{cases},  & \rm{for~Case~3'} \\
    \rho_{{\rm so}, \nu}^{\texttt{NL}}  ,   &   \rm{for~Case~4'}  \\
    1  ,   &   \rm{for~Case~5'}
\end{cases}. \tag{14}
\end{align}
\hrulefill
\end{figure*}

By Lemma \ref{lem_1}, it is possible to find the optimal solution to (P1) based on the Lagrange duality method.
However, even if the Lagrange duality method can be used, it is still difficult to obtain the solution to (P1) because the Lagrangian is a nonconvex function of $\{ \rho_{\nu} \}$.
To overcome this difficulty, we take the following approach: we first classify the conditions of the channel power gain $h_{\nu}$ into the five mutually exclusive cases,
and then, we derive the optimal solution in each case.
Taking this approach, we have the following result.
\begin{theorem}
\label{thm_1}
The solution to (P1) is given by (\ref{rho_NL}) (shown at the top of this page),
where $\rho_{{\rm o}, \nu}^{\texttt{NL}} $ is the root of the following equation:
\begin{align}
\label{rho_opt}
& \lambda^{\texttt{NL}} \Psi_{\nu} ( P, x  ) \left(  1 - \Psi_{\nu} ( P, x  ) \right) = \frac{1 - \Omega }{  P_{s} T a \left(  (1 - x ) h_{\nu} P + \sigma^{2} \right) }  \tag{10}
\end{align}
over $ x \in \left(  \frac{b}{h_{\nu} P} , 1 \right) $.
Also, the Cases 1--5 are given by (\ref{rho_NL_cases}) (shown at the top of this page),
where $ f ( x ) =   \frac{ e^{ - a ( x P - b ) }  }{  \left( 1 + e^{ - a ( x P - b ) } \right)^{2} }  $, $ g ( x ) =  \frac{1 - \Omega }{ P_{s} T a ( x P + \sigma^{2} ) } $, and $ \gamma( x ) =  \frac{1 - \Omega }{ P_{s} T a \left( \max \{ x P - b , 0 \} + \sigma^{2} \right) } $.
The constant $ \lambda^{\texttt{NL}} > 0 $ is chosen to satisfy $\mathbb{E} [   Q_{\nu}^{\texttt{NL}} ( P , \rho_{\nu}^{\texttt{NL}} ) ]  = Q$.
\end{theorem}
\begin{IEEEproof}
See Appendix \ref{proof_thm_1}.
\end{IEEEproof}

From Theorem \ref{thm_1}, the optimal solution to (P1) can be computed efficiently.
Specifically, at each fading state $\nu$, one can determine the optimal power splitting ratio $ \rho_{\nu}^{\texttt{NL}} $ according to (\ref{rho_NL})
by simply checking which case the channel power gain $h_{\nu}$ falls into (through the functions $f(\cdot)$, $g(\cdot)$, and $\gamma(\cdot)$).
In the proposed solution, only the positive scalars $ \lambda^{\texttt{NL}}$ and $ \{ \rho_{{\rm o}, \nu}^{\texttt{NL}} \} $ need to be computed numerically.
The value of $\lambda^{\texttt{NL}}$ can be determined efficiently via the subgradient method \cite{Yu},
of which the complexity is given by $\mathcal{O} ( 1^{2} ) = \mathcal{O} ( 1 )$ \cite{Boyd}.
Also, the value of $ \rho_{{\rm o} , \nu}^{\texttt{NL}} $ can be determined efficiently via the one-dimensional searching, e.g., the bisection method,
of which complexity is at most $ \mathcal{O} ( K ) $ \cite{Boyd}, where the parameter $ K > 1$ is inversely proportional to the tolerance.
Overall, the computational complexity to solve (P1) is given by $\mathcal{O} (K N)$.
Note that the complexity of the proposed solution is linear in the number $N$ of fading blocks,
and thus, it is much lower than the exponential complexity $ \mathcal{O} ( M^{N} ) $ required by the exhaustive searching, where $M \geq 2$ is the parameter related to the searching resolution.
Since $N$ is large in practice, the complexity reduction by Theorem \ref{thm_1} is indeed significant.

\subsection{Suboptimal Solution to (P1)}

The computational complexity of the optimal solution to (P1) in Theorem \ref{thm_1} is low.
Unfortunately, the complexity of the solution in Theorem \ref{thm_1} might not be low enough for certain practical applications,
because it is proportional to both $N$ and $K$.
For example, when $K \geq N$, the complexity is higher than $\mathcal{O} ( N^{2})$.
In the practical fading scenario and from the practical implementation perspective, it is very desirable to achieve the complexity proportional only to the number $N$ of the fading blocks,
i.e., in the order of $\mathcal{O} (N)$ \cite{Yu}.
However, in (P1), it is very challenging to reduce the complexity while guaranteeing the optimality.
To address this issue, in this subsection, we propose a suboptimal solution to (P1) in closed form (up to the Lagrange multiplier $\lambda^{\texttt{NL}}$),
which turns out to be asymptotically optimal in the low SNR region.

\setcounter{equation}{11}

The fundamental idea is to maximize the lower bound of the average rate rather than the actual average rate.
To this end, in the following, we first derive a lower bound of the average rate.
\begin{lemma}
\label{lem_lb}
The objective function of the problem (P1) is lower bounded by
\begin{align}
\label{R_lb}
\mathbb{E} \left[  R_{\nu} ( P , \rho_{\nu}  )  \right]  \geq \mathbb{E} \left[  (1 - \rho_{\nu} ) R_{\nu} ( P , 0  ) \right]  .
\end{align}
\end{lemma}
\begin{IEEEproof}
Since $ R_{\nu} ( P , \rho_{\nu}  ) $ is concave in $ (1 - \rho_{\nu}) $ \cite{Zhou13}, it follows that $ R_{\nu} ( P , \rho_{\nu}  ) \geq (1 - \rho_{\nu} ) R_{\nu} ( P , 0  ) $, $\forall \nu$ \cite{Boyd}.
\end{IEEEproof}

The lower bound of (\ref{R_lb}) becomes tight when $\rho_{\nu} \in \{ 0,1 \}$, $\forall \nu$.
Replacing the objective function of (P1) by the lower bound of (\ref{R_lb}),
we can formulate the following problem:
\begin{subequations}
\label{P1_}
\begin{align}
\label{P1_obj_}
{\rm (P1'):}  \quad  \underset{ 0 \leq \rho_{\nu} \leq 1,  \forall \nu }{\texttt{max}} & \quad \mathbb{E} \left[ (1 - \rho_{\nu} ) R_{\nu} ( P , 0  )  \right]   \\
\label{P1_const_1_}
\texttt{s.t.}  &  \quad  \mathbb{E} \left[  Q_{\nu}^{\texttt{NL}} (  P , \rho_{\nu}  )  \right] \geq Q.
\end{align}
\end{subequations}

\setcounter{equation}{14}

In the following, the solution to (P1$'$) is derived.
\begin{theorem}
\label{thm_2}
The solution to (P1$'$) is given by (\ref{rho_NL_so}) (shown at the top of the previous page),
where the Cases 1$'$--5$'$ are given by
\begin{align}
\label{rho_NL_so_cases}
& z ( h_{\nu} ) \geq  \tfrac{\lambda^{\texttt{NL}}}{4}    && \quad {\rm for ~ Case ~ 1'}  \nonumber \\
& \max \left\{ \lambda^{\texttt{NL}} f ( h_{\nu} ), \lambda^{\texttt{NL}} f ( 0 )  \right\} < z ( h_{\nu} )  < \tfrac{\lambda^{\texttt{NL}}}{4}  && \quad {\rm for ~ Case ~ 2'}  \nonumber\\
& \lambda^{\texttt{NL}} f ( 0 )  \leq  z ( h_{\nu} )  \leq \lambda^{\texttt{NL}} f ( h_{\nu} )  && \quad {\rm for ~ Case ~ 3'}  \nonumber\\
& \lambda^{\texttt{NL}} f ( h_{\nu} )  <  z ( h_{\nu} )  < \lambda^{\texttt{NL}} f ( 0 )  && \quad {\rm for ~ Case ~ 4'} \nonumber \\
& z ( h_{\nu} )  \leq   \min \left\{ \lambda^{\texttt{NL}} f ( h_{\nu} ), \lambda^{\texttt{NL}} f ( 0 )  \right\}  && \quad {\rm for ~ Case ~ 5'}  .
\end{align}
In (\ref{rho_NL_so_cases}), $ f ( x ) =   \frac{ e^{ - a ( x P - b ) }  }{  \left( 1 + e^{ - a ( x P - b ) } \right)^{2} }  $ and $ z ( x ) =  \frac{1 - \Omega }{ P_{s} T a x P } \log_{2} \left(  1 + \frac{x P}{\sigma^{2}} \right) $.
Also, $\rho_{{\rm so}, \nu}^{\texttt{NL}}$ is given by
\begin{align}
\label{rho_subopt}
\rho_{{\rm so}, \nu}^{\texttt{NL}}  =  \frac{1}{h_{\nu} P} \left( - \frac{1}{a}  \ln \left(  \frac{ 2 }{  1 + \sqrt{ 1 - \frac{ 4 z ( h_{\nu} ) }{ \lambda^{\texttt{NL}} } }  } - 1 \right) + b \right).
\end{align}
The constant $ \lambda^{\texttt{NL}} > 0 $ is chosen to satisfy $\mathbb{E} [   Q_{\nu}^{\texttt{NL}} ( P , \rho_{\nu}^{\texttt{NL}'} ) ]  = Q$.
\end{theorem}
\begin{IEEEproof}
The result can be proved by following the similar procedures in Appendix \ref{proof_thm_1} and replacing the objective function $ \mathbb{E} \left[  R_{\nu} ( P , \rho_{\nu}  )  \right]  $ by the lower bound $\mathbb{E} \left[ (1 - \rho_{\nu} ) R_{\nu} ( P , 0  )  \right]$, where $\rho_{{\rm so}, \nu}^{\texttt{NL}}$ in (\ref{rho_subopt}) is given by the solution of the equation $ \Psi_{\nu} ( P, x  ) =  \frac{1}{2} + \sqrt{ \frac{1}{4} - \frac{z(h_{\nu})}{\lambda^{\texttt{NL}}} } $ over $ x \in \Big( \frac{b}{h_{\nu} P},  1 \Big) $.
\end{IEEEproof}

Note that the solution in Theorem \ref{thm_2} can be computed more efficiently than the optimal solution in Theorem \ref{thm_1}.
Specifically, only a single positive scalar $ \lambda^{\texttt{NL}} $ needs to be computed numerically.
Consequently, the complexity of the solution in Theorem \ref{thm_2} is given by $\mathcal{O} (N)$, which is proportional only to $N$, and thus, is much lower than the complexity $\mathcal{O} (K N)$ of the optimal solution.
But, the solution in Theorem \ref{thm_2} is generally suboptimal to (P1) because the lower bound of (\ref{R_lb}) is used rather than the exact value.
Interestingly and fortunately, we can show that the solution in Theorem \ref{thm_2} is asymptotically optimal in the low SNR regime.
\begin{lemma}
\label{lem_optimality}
When $ \frac{ P }{ \sigma^{2} } \rightarrow 0 $, the solution in Theorem \ref{thm_2} is optimal to (P1).
\end{lemma}
\begin{IEEEproof}
It follows that $  R_{\nu} ( P , \rho_{\nu}  ) \rightarrow \frac{ (1 - \rho_{\nu}) h_{\nu} P }{ \sigma^{2} \ln 2 } $ and $ (1 - \rho_{\nu} ) R_{\nu} ( P , 0  ) \rightarrow \frac{ (1 - \rho_{\nu}) h_{\nu} P }{ \sigma^{2} \ln 2 } $, as $ \frac{ P }{ \sigma^{2} } \rightarrow 0 $ \cite{Tse}.
Thus, when $ \frac{ P }{ \sigma^{2} } \rightarrow 0 $, the gap between the actual objective value of (P1) and the lower bound of (\ref{R_lb}) approaches zero, i.e., $ R_{\nu} ( P , \rho_{\nu}  ) - (1 - \rho_{\nu} ) R_{\nu} ( P , 0  ) \rightarrow 0 $, implying that the solution in Theorem \ref{thm_2} is optimal to (P1).
\end{IEEEproof}

Although we can mathematically show the optimality of the solution in Theorem \ref{thm_2} only in the low SNR range,
the numerical results in Section \ref{numerical_result} will demonstrate that the performance of this solution is essentially the same as the optimal performance even in the moderate to high SNR range.

\subsection{Comparisons to Linear EH}
\label{sec_P_EH_comp}

In this subsection, to obtain new and useful insights, we compare the proposed scheme to the existing dynamic power splitting scheme for the linear EH \cite[Proposition 4.1]{Liu_dps}:
\begin{align}
\label{rho_L}
\rho_{\nu}^{\texttt{L}}  =  \begin{cases} \rho_{{\rm o}, \nu}^{\texttt{L}}  ,  &  {\rm if}  ~  h_{\nu}  >  x_{1}^{\texttt{L}}  \\ 0, & {\rm otherwise}  \end{cases}
\end{align}
where $ \rho_{{\rm o}, \nu}^{\texttt{L}} = 1 - \frac{1}{h_{\nu} P} \left( \frac{1}{ \zeta \lambda^{\texttt{L}} } - \sigma^{2} \right) $.
Also, $x_{1}^{\texttt{L}} = \frac{1}{ P }  \left( \frac{1}{ \zeta \lambda^{\texttt{L}} } - \sigma^{2} \right)$.
The constant $ 0 < \lambda^{\texttt{L}} \leq \frac{1}{\zeta \sigma^{2}} $ is determined such that $\mathbb{E} [   Q_{\nu}^{\texttt{L}} ( P , \rho_{\nu}^{\texttt{L}} ) ]  = Q$.

\begin{figure}
    \centering
    \subfigure[Dynamic power splitting for the linear EH, eq. (\ref{rho_L}) \cite{Liu_dps}]
    {
        \includegraphics[width=0.45\textwidth]{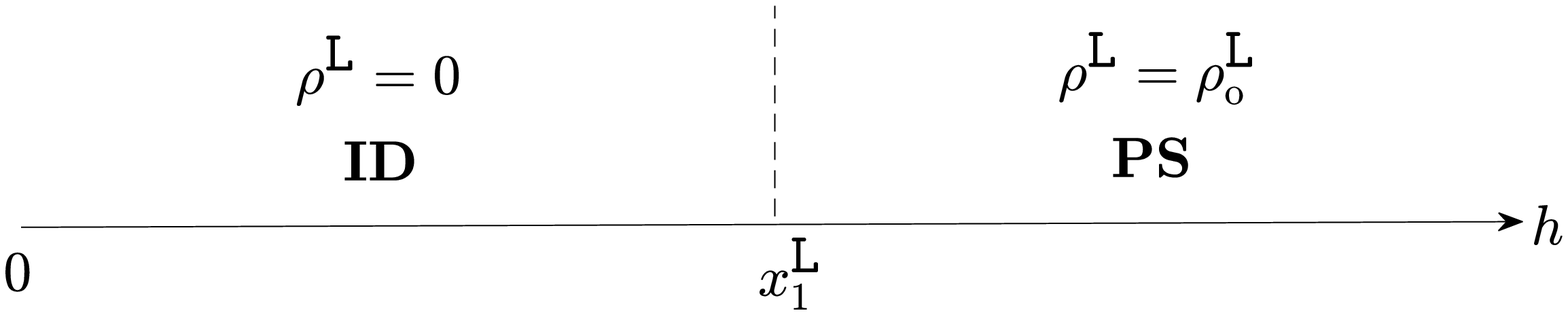}
        \label{fig1_a}
    }
    \subfigure[Dynamic power splitting for the nonlinear EH, eq. (\ref{rho_NL}) (proposed)]
    {
        \includegraphics[width=0.45\textwidth]{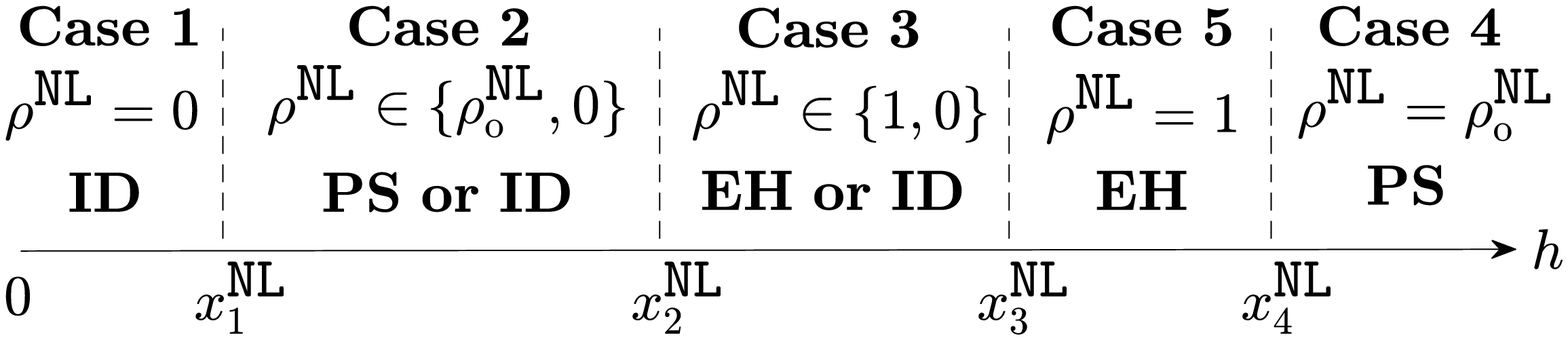}
        \label{fig1_c}
    }
    \caption{Comparison of the proposed and existing schemes for the case of CSIR.}
    \label{fig1}
\end{figure}

In Fig. \ref{fig1}, we illustrate the proposed and existing schemes.
In this figure, for notational simplicity, we focus on a particular fading state, and thus, we drop the index $\nu$ from the relevant expressions.
Also, in Fig. \ref{fig1}, the term ``PS'' means the power splitting, i.e., both ID and EH.
In Fig. \ref{fig1_c}, $x_{1}^{\texttt{NL}}$, $( x_{2}^{\texttt{NL}} , x_{4}^{\texttt{NL}} )$, and $x_{3}^{\texttt{NL}}$ denote the roots of the equations
$ \gamma(x) = \frac{1}{4}$, $ \lambda_{\texttt{NL}} f(x) = g(0)$, and $ \lambda_{\texttt{NL}} f(0) = g(x)$, respectively.
For the illustration purpose, we assume that $ x_{j}^{\texttt{NL}} < x_{j+1}^{\texttt{NL}} $, $j = 1,2,3$.

To obtain the direct and useful insights that are related to the distance, 
we consider a practical distance-based channel model of \cite{Huang15}:
$h = 1 - \exp\left( - \frac{a_{t} a_{r}}{ (c/f_{c})^{2} d^{2} } \right)$,
where $a_{t}$ is the aperture of the transmit antenna; $a_{r}$ is the aperture of the receive antenna; $c$ is the speed of light; $f_{c}$ is the carrier frequency; and $d$ is the distance between the transmitter and the receiver.
Then, comparing the proposed and existing schemes, we can obtain the following insights:
\begin{itemize}
\item From Figs. \ref{fig1_a} and \ref{fig1_c}, and the results of (\ref{rho_NL}) and (\ref{rho_L}), one can see that the optimal dynamic power splitting schemes for the linear EH and the nonlinear EH are fundamentally different.
    For example, for the case of linear EH, the received power is used for the two different purposes: only for ID with $\rho^{\texttt{L}} = 0$ or for both ID and EH with $0 < \rho^{\texttt{L}} < 1$. Thus, in the linear EH, there are \textit{two} different regions for the optimal dynamic power splitting. On the other hand, for the case of nonlinear EH, the received power is used for the three different purposes: only for ID with $\rho^{\texttt{NL}} = 0$, only for EH with $\rho^{\texttt{NL}} = 1$, or for both ID and EH with $0 < \rho^{\texttt{NL}} < 1$. Consequently, in the nonlinear EH, there are up to \textit{five} different regions for the optimal dynamic power splitting.
    Also, for the case of linear EH, the optimal power splitting ratio $\rho^{\texttt{NL}}$ increases as the distance $d$ decreases (or the channel gain $h$ increases), and thus, the power used for EH increases.
    On the other hand, for the case of nonlinear EH, the optimal power splitting ratio $\rho^{\texttt{NL}}$ (and thus, the power used for EH) increases, and then, decreases as the distance $d$ decreases.
\end{itemize}

\subsection{Extension to the Case of Partial CSI at the Transmitter}
In the previous subsections, we studied the case of no CSI at the transmitter. 
In this subsection, we extend our analysis to the case of the partial CSI at the transmitter where the long-term statistics such as the mean or variance of the channel distribution is known at the transmitter.
In this case, the problem to achieve the optimal R-E tradeoff performance can be formulated as the problem (P1), but additionally optimizing the transmit power $P$ as follows:
$ \underset{ \substack{0 \leq \rho_{\nu} \leq 1,  \forall \nu, \\ 0 \leq P \leq P_{\rm th}} }{\texttt{max}}  ~ \mathbb{E} \left[  R_{\nu} ( P , \rho_{\nu}  )  \right]  ~ \texttt{s.t.}  ~ \mathbb{E} \left[  Q_{\nu}^{\texttt{NL}} (  P , \rho_{\nu}  )  \right] \geq Q $, where $P_{\rm th}$ denotes the threshold for the transmit power.
This problem can be solved by extending our analytical approaches to solve (P1).
Specifically, following the similar procedures in Appendix \ref{proof_lems_1_2}, 
it can be shown that the time-sharing condition holds. 
Thus, using the Lagrange duality method and following the similar procedures in Appendix \ref{proof_thm_1},
it can be shown that the solution is given by $ (  \rho_{\nu}^{*} ( P^{*} ) ,  P^{*} ) $,
where $   \rho_{\nu}^{*} ( P^{*} ) $ is given by the solution derived in Theorem \ref{thm_1} with $P$ replaced by $P^{*}$.
Also, $P^{*}$ is the solution to the following problem: $\underset{ 0 \leq P \leq P_{\rm th} }{\texttt{max}} ~ \big\{ \mathbb{E} \left[  R_{\nu} ( P , \rho_{\nu}^{*} (P)  )  \right]  + \lambda^{\texttt{NL}} \mathbb{E} \left[  Q_{\nu}^{\texttt{NL}} (  P , \rho_{\nu}^{*} (P)  )  \right] \big\}$, which can be determined via one-dimensional searching.
The constant $\lambda^{\texttt{NL}} > 0$ is determined such that $\mathbb{E} \left[  Q_{\nu}^{\texttt{NL}} (  P^{*} , \rho_{\nu}^{*} (P^{*})  )  \right] = Q$.
Overall, the optimal dynamic power splitting scheme with partial CSI at the transmitter has the complexity of $\mathcal{O}(K^{2} N)$.

\section{Dynamic Power Splitting for Nonlinear EH with CSI}
\label{sec_mode_PA}

In this section, we consider the case of CSI. We first derive the optimal solution to (P2).
Then a suboptimal solution to (P2) is proposed with low complexity.
Finally, we compare the proposed and existing schemes.

\subsection{Optimal Solution to (P2)}

In order to solve (P2), in the following, we first define the time-sharing condition for (P2).
\begin{definition}[\textit{\cite[Definition 1]{Yu}}]
Let $\{ P_{x, \nu}^{*} , \rho_{ x, \nu}^{*} \} $ and $\{ P_{y, \nu}^{*} , \rho_{y, \nu}^{*} \}$ denote the optimal solutions to the problem (P2) with $ ( Q, P_{\rm avg} ) = ( Q_{x} , \overline{P}_{x} ) $ and $( Q, P_{\rm avg} ) = (Q_{y}, \overline{P}_{y} )$, respectively. Then the problem (P2) is said to satisfy the \textit{time-sharing} condition (or time-sharing property) if the following condition holds: there exists a feasible point $\{ P_{z, \nu} , \rho_{z, \nu} \}$ satisfying $\mathbb{E} [  Q_{\nu}^{\texttt{NL}} (  P_{z, \nu}, \rho_{z, \nu}   )  ] \geq \theta Q_{x} + (1 - \theta) Q_{y}$, $ \mathbb{E} [ P_{z, \nu} ] \leq  \theta \overline{P}_{x}  + (1 - \theta) \overline{P}_{y} $, and $ \mathbb{E} [ R_{\nu} ( P_{z, \nu} , \rho_{z, \nu}) ] \geq  \theta \mathbb{E} [ R_{\nu} ( P_{x, \nu}^{*} , \rho_{x, \nu}^{*}) ] + (1 - \theta) \mathbb{E} [ R_{\nu} ( P_{y, \nu}^{*} , \rho_{y, \nu}^{*}) ] $ for any $( Q_{x} , \overline{P}_{x} )$, $( Q_{y} , \overline{P}_{y} )$, and $0 \leq \theta \leq 1$.
\null \hfill $\blacksquare$
\end{definition}

Now, we establish the time-sharing property of (P2).
\begin{lemma}
\label{lem_2}
In the problem (P2), the time-sharing condition is satisfied.
\end{lemma}
\begin{IEEEproof}
See Appendix \ref{proof_lems_1_2}.
\end{IEEEproof}

\begin{figure*}[t]
\begin{align}
\label{P_EH_NL}
P_{{\rm EH}, \nu}^{\texttt{NL}}  = \begin{cases}  \left[ P_{{\rm EH}, \nu}^{A} \right]_{0}^{ P_{\max} }   , &  {\rm if} ~ h_{\nu} \leq  \mu^{\texttt{NL}} \sigma^{2} \\ \begin{cases} \left[ P_{{\rm EH},\nu}^{A} \right]_{0}^{ P_{{\rm th}, \nu} }  , &  {\rm if} ~ \log_{2} \left( \frac{h_{\nu}}{\lambda^{\texttt{NL}} \sigma^{2}} \right) + \frac{\lambda^{\texttt{NL}} \sigma^{2}}{h_{\nu}} - 1  + \lambda^{\texttt{NL}} Q_{\nu}^{\texttt{NL}} \left( \left[ P_{{\rm EH}, \nu}^{A} \right]_{0}^{ P_{{\rm th}, \nu} } , 1 \right)  - \mu^{\texttt{NL}} \cdot \left[ P_{{\rm EH}, \nu}^{A} \right]_{0}^{ P_{{\rm th}, \nu} }   \\ & \quad  > R_{\nu} \left( P_{\max} - \left[ P_{{\rm EH}, \nu}^{B} \right]_{ P_{{\rm th}, \nu}  }^{ P_{\max} } , 0 \right) + \lambda^{\texttt{NL}} Q_{\nu}^{\texttt{NL}} \left( \left[ P_{{\rm EH}, \nu}^{B} \right]_{  P_{{\rm th}, \nu} }^{ P_{\max} } , 1 \right) - \mu^{\texttt{NL}} P_{\max}   \\ \left[ P_{{\rm EH},\nu}^{B} \right]_{  P_{{\rm th}, \nu} }^{ P_{\max} } , &  {\rm otherwise} \end{cases} , & {\rm if} ~ h_{\nu} >  \mu^{\texttt{NL}} \sigma^{2}  \end{cases}. \tag{20}
\end{align}
\hrulefill
\begin{align}
\label{p_EH_NL_A}
\left[ P_{{\rm EH}, \nu}^{A} \right]_{P_{\rm low}}^{ P_{\rm up} } = \begin{cases}  P_{\rm low}  ,   &   {\rm for~Case~A . 1}   \\
    \begin{cases}  P_{{\rm o}, \nu}^{ A}  ,  &  {\rm if} ~  \lambda^{\texttt{NL}} Q_{\nu}^{\texttt{NL}} ( P_{{\rm o}, \nu}^{ A} , 1 ) - \mu^{\texttt{NL}} P_{{\rm o}, \nu}^{ A} > \lambda^{\texttt{NL}} Q_{\nu}^{\texttt{NL}} ( P_{\rm low} , 1 ) - \mu^{\texttt{NL}} P_{\rm low} \\ P_{\rm low}, & {\rm otherwise}  \end{cases},  &  {\rm for~Case~A . 2} \\
    \begin{cases}  P_{\rm up} ,  &  {\rm if}  ~ \lambda^{\texttt{NL}} Q_{\nu}^{\texttt{NL}} ( P_{\rm up} , 1 ) - \mu^{\texttt{NL}} P_{\rm up} > \lambda^{\texttt{NL}} Q_{\nu}^{\texttt{NL}} ( P_{\rm low} , 1 ) - \mu^{\texttt{NL}} P_{\rm low} \\ P_{\rm low}, & {\rm otherwise}  \end{cases},  & {\rm for~Case~A . 3} \\
    P_{{\rm o}, \nu}^{ A}   ,   &   {\rm for~Case~A . 4}  \\
    P_{\rm up}   ,   &   {\rm for~Case~A . 5}
\end{cases}. \tag{21}
\end{align}
\hrulefill
\begin{align}
\label{p_EH_NL_A_cases}
& \mu^{\texttt{NL}} Z ( h_{\nu} ) \geq  \tfrac{\lambda^{\texttt{NL}}}{4}    && \quad {\rm for ~ Case ~ A . 1}  \nonumber \\
& \max \left\{ \lambda^{\texttt{NL}} F_{\rm up} ( h_{\nu} ), \lambda^{\texttt{NL}} F_{\rm low} ( h_{\nu} )  \right\} < \mu^{\texttt{NL}} Z ( h_{\nu} )  < \tfrac{\lambda^{\texttt{NL}}}{4}  && \quad {\rm for ~ Case ~ A . 2} \nonumber \\
& \lambda^{\texttt{NL}} F_{\rm low} ( h_{\nu} )  \leq  \mu^{\texttt{NL}} Z ( h_{\nu} )  \leq \lambda^{\texttt{NL}} F_{\rm up} ( h_{\nu} )  && \quad {\rm for ~ Case ~ A . 3}  \nonumber\\
& \lambda^{\texttt{NL}} F_{\rm up} ( h_{\nu} )  \leq  \mu^{\texttt{NL}} Z ( h_{\nu} )  \leq \lambda^{\texttt{NL}} F_{\rm low} ( h_{\nu} )  && \quad {\rm for ~ Case ~ A . 4}  \nonumber\\
& \mu^{\texttt{NL}} Z ( h_{\nu} )  \leq   \min \left\{ \lambda^{\texttt{NL}} F_{\rm up} ( h_{\nu} ), \lambda^{\texttt{NL}} F_{\rm low} ( h_{\nu} )  \right\}  && \quad {\rm for ~ Case ~ A . 5}. \tag{22}
\end{align}
\hrulefill
\begin{align}
\label{p_EH_NL_B}
\left[ P_{{\rm EH}, \nu}^{B} \right]_{P_{\rm low}}^{ P_{\rm up} }  = \begin{cases}  P_{\rm low}  ,   &   {\rm for~Case~B . 1}   \\
    \begin{cases}  P_{{\rm o}, \nu}^{ B}  ,  &  {\rm if} ~  R_{\nu} ( P_{\rm up} - P_{{\rm o}, \nu}^{ B} , 0 ) + \lambda^{\texttt{NL}} Q_{\nu}^{\texttt{NL}} ( P_{{\rm o}, \nu}^{ B} , 1 )  > R_{\nu} ( P_{\rm up} - P_{\rm low} , 0 ) + \lambda^{\texttt{NL}} Q_{\nu}^{\texttt{NL}} ( P_{\rm low} , 1 ) \\ P_{\rm low}, & {\rm otherwise}  \end{cases},  &  {\rm for~Case~B . 2} \\
    \begin{cases}  P_{\rm up} ,  &  {\rm if}  ~  \lambda^{\texttt{NL}} Q_{\nu}^{\texttt{NL}} ( P_{\rm up} , 1 ) > R_{\nu} ( P_{\rm up} - P_{\rm low} , 0 ) + \lambda^{\texttt{NL}} Q_{\nu}^{\texttt{NL}} ( P_{\rm low} , 1 )  \\ P_{\rm low}, & {\rm otherwise}  \end{cases},  & {\rm for~Case~B . 3} \\
    P_{{\rm o}, \nu}^{ B}   ,   &   {\rm for~Case~B . 4}  \\
    P_{\rm up}   ,   &   {\rm for~Case~B . 5}
\end{cases}. \tag{23}
\end{align}
\hrulefill
\begin{align}
\label{p_EH_NL_B_cases}
& \Gamma ( h_{\nu} )  \geq \tfrac{\lambda^{\texttt{NL}}}{4}   && \quad {\rm for ~ Case ~ B . 1}  \nonumber\\
& \lambda^{\texttt{NL}} F_{\rm low} ( h_{\nu} ) \leq G_{\rm low} ( h_{\nu} ) , ~   \lambda^{\texttt{NL}} F_{\rm up} ( h_{\nu} )  <  G_{\rm up} ( h_{\nu} ) ,  ~ {\rm and} ~  \Gamma ( h_{\nu} ) < \tfrac{\lambda^{\texttt{NL}}}{4}  && \quad {\rm for ~ Case ~ B . 2} \nonumber\\
& \lambda^{\texttt{NL}} F_{\rm low} ( h_{\nu} ) \leq G_{\rm low} ( h_{\nu} ) , ~  \lambda^{\texttt{NL}} F_{\rm up} ( h_{\nu} )  \geq  G_{\rm up} ( h_{\nu} ) ,  ~ {\rm and} ~  \Gamma ( h_{\nu} ) < \tfrac{\lambda^{\texttt{NL}}}{4}  && \quad {\rm for ~ Case ~ B . 3} \nonumber\\
& \lambda^{\texttt{NL}} F_{\rm low} ( h_{\nu} ) > G_{\rm low} ( h_{\nu} ) , ~   \lambda^{\texttt{NL}} F_{\rm up} ( h_{\nu} )  <  G_{\rm up} ( h_{\nu} ) ,  ~ {\rm and} ~  \Gamma ( h_{\nu} ) <  \tfrac{\lambda^{\texttt{NL}}}{4}  && \quad {\rm for ~ Case ~ B . 4} \nonumber\\
& \lambda^{\texttt{NL}} F_{\rm low} ( h_{\nu} ) > G_{\rm low} ( h_{\nu} ) , ~ \lambda^{\texttt{NL}} F_{\rm up} ( h_{\nu} ) \geq  G_{\rm up} ( h_{\nu} ) ,  ~ {\rm and} ~  \Gamma ( h_{\nu} ) < \tfrac{\lambda^{\texttt{NL}}}{4} && \quad {\rm for ~ Case ~ B . 5}. \tag{24}
\end{align}
\hrulefill
\end{figure*}

From Lemma \ref{lem_2}, the optimal solution to (P2) can be obtained by using the Lagrange duality method, similar to (P1).
However, even when the Lagrange duality method is used, the problem (P2) is still very difficult to solve (much more difficult than (P1)),
because the variables $\{ \rho_{\nu} \}$ and $\{ P_{\nu} \}$ are coupled.
To overcome the difficulty, we take the following approach: we first convert the problem (P2) into a more tractable form, and then, we derive the solution by solving the converted problem.
Taking this approach, we have the following result.
\begin{theorem}
\label{thm_3}
The solution to (P2) is given by
\begin{subequations}
\label{P2_sol}
\begin{align}
\label{P2_sol_P}
P_{\nu}^{\texttt{NL}} & = P_{{\rm EH}, \nu}^{\texttt{NL}} + P_{{\rm ID}, \nu}^{\texttt{NL}}, \tag{18a} \\
\label{P2_sol_rho}
\rho_{\nu}^{\texttt{NL}} & = \begin{cases}  0 , & {\rm if ~} P_{{\rm ID}, \nu}^{\texttt{NL}} = P_{{\rm EH}, \nu}^{\texttt{NL}} = 0 \\ \frac{P_{{\rm EH}, \nu}^{\texttt{NL}}}{P_{{\rm ID}, \nu}^{\texttt{NL}} + P_{{\rm EH}, \nu}^{\texttt{NL}}}, & {\rm otherwise} \end{cases} . \tag{18b}
\end{align}
\end{subequations}
In (\ref{P2_sol}), $ P_{{\rm ID}, \nu}^{\texttt{NL}} $ is given by
\begin{align}
\label{P_ID_NL}
P_{{\rm ID}, \nu}^{\texttt{NL}} &  =  \begin{cases} 0 , &  {\rm if} ~ h_{\nu} \leq  \mu^{\texttt{NL}} \sigma^{2}  \\  \min \left\{ \frac{1}{\mu^{\texttt{NL}} } - \frac{\sigma^{2}}{h_{\nu}} , P_{\max}  - P_{{\rm EH}, \nu}^{\texttt{NL}}  \right\}  , &   {\rm if} ~ h_{\nu} >  \mu^{\texttt{NL}} \sigma^{2}   \end{cases}. \tag{19}
\end{align}
Also, $ P_{{\rm EH}, \nu}^{\texttt{NL}} $ is given by (\ref{P_EH_NL}) (shown at the top of this page),
where $ P_{{\rm th}, \nu} =  \max \Big\{ P_{\max} - \frac{1}{\mu^{\texttt{NL}} } +  \frac{ \sigma^{2} }{ h_{\nu} }, 0  \Big\} $.
In (\ref{P_EH_NL}), $ \left[ P_{{\rm EH}, \nu}^{A} \right]_{P_{\rm low}}^{ P_{\rm up} } $ is given by (\ref{p_EH_NL_A}) (shown at the top of this page),
where the Cases A.1--A.5 are given by (\ref{p_EH_NL_A_cases}) (shown at the top of this page).
Also, $\left[ P_{{\rm EH}, \nu}^{B} \right]_{P_{\rm low}}^{ P_{\rm up} }$ is given by (\ref{p_EH_NL_B}) (shown at the top of this page),
where the Cases B.1--B.5 are given by (\ref{p_EH_NL_B_cases}) (shown at the top of this page).
In (\ref{p_EH_NL_A_cases}) and (\ref{p_EH_NL_B_cases}),
$ F_{\rm low/up} ( x  ) =   \frac{ e^{ - a ( x P_{\rm low/up} - b ) }  }{  \left( 1 + e^{ - a ( x P_{\rm low/up} - b ) } \right)^{2} }  $, $ G_{\rm low/up} ( x  ) =  \frac{1 - \Omega }{ P_{s} T a ( x ( P_{\max} - P_{\rm low/up} ) + \sigma^{2} ) } $, $ Z (x) =   \frac{1 - \Omega  }{ P_{s} T a x } $, and $ \Gamma( x ) =  \frac{1 - \Omega }{ P_{s} T a \left( \max \{ x P_{\max} - b , 0 \} + \sigma^{2} \right) } $.
In (\ref{p_EH_NL_A}), $ P_{{\rm o}, \nu}^{ A} $ is given by
\begin{align}
\label{P_A_opt}
P_{{\rm o}, \nu}^{ A}  =  \frac{1}{h_{\nu}} \left( - \frac{1}{a}  \ln \left(  \frac{ 2 }{  1 + \sqrt{ 1 - \frac{ 4 \mu^{\texttt{NL}} Z ( h_{\nu} ) }{ \lambda^{\texttt{NL}} } }  } - 1 \right) + b \right). \tag{25}
\end{align}
In (\ref{p_EH_NL_B}), $ P_{{\rm o}, \nu}^{ B} $ is the root of the following equation:
\begin{align}
\label{P_B_opt}
\lambda^{\texttt{NL}} \Psi_{\nu} ( x, 1  ) \left(  1 - \Psi_{\nu} ( x, 1  ) \right) = \frac{1 - \Omega }{  P_{s} T a \left( h_{\nu} ( P_{\max} - x ) + \sigma^{2} \right) } \tag{26}
\end{align}
over $ x \in \left(  \frac{b}{h_{\nu}} , P_{\rm up} \right) $.
The constants $\lambda^{\texttt{NL}} > 0$ and $\mu^{\texttt{NL}} > 0$ are chosen to satisfy $\mathbb{E} \left[  Q_{\nu}^{\texttt{NL}} (  P_{\nu}^{\texttt{NL}} , \rho_{\nu}^{\texttt{NL}}  )  \right] = Q$ and $\mathbb{E} \left[  P_{\nu}^{\texttt{NL}}  \right]  = P_{\rm avg}$.
\end{theorem}
\begin{IEEEproof}
See Appendix \ref{proof_thm_3}.
\end{IEEEproof}

The result of Theorem \ref{thm_3} can be intuitively interpreted as follows.
The terms $P_{{\rm ID}, \nu}^{\texttt{NL}}$ of (\ref{P_ID_NL}) and $P_{{\rm EH}, \nu}^{\texttt{NL}}$ of (\ref{P_EH_NL}) can be considered as the optimal power allocations for ID and EH, respectively.
According to (\ref{P2_sol_P}), the optimal transmit power is given by the total amount of power assigned to both ID and EH, i.e., $P_{\nu}^{\texttt{NL}} = P_{{\rm ID}, \nu}^{\texttt{NL}} + P_{{\rm EH}, \nu}^{\texttt{NL}}$.
Also, according to (\ref{P2_sol_rho}),
the optimal power splitting ratio is set to the fraction of power assigned to EH, i.e., $  \rho_{\nu}^{\texttt{NL}} = \frac{P_{{\rm EH}, \nu}^{\texttt{NL}}}{P_{{\rm ID}, \nu}^{\texttt{NL}} + P_{{\rm EH}, \nu}^{\texttt{NL}}} $, and it is set to zero if the optimal transmit power is zero.


From Theorem \ref{thm_3}, the optimal solution to (P2) can be computed efficiently.
In the proposed solution, the positive scalars $\lambda^{\texttt{NL}}$, $\mu^{\texttt{NL}}$, and $\{ P_{{\rm o}, \nu}^{ B} \}$ need to be computed numerically.
The values of $\lambda^{\texttt{NL}}$ and $\mu^{\texttt{NL}}$ can be determined efficiently via the sub-gradient method \cite{Yu}.
Also, the value of $ P_{{\rm o}, \nu}^{ B} $ can be determined efficiently via the one-dimensional searching.
Overall, the computational complexity to solve (P2) is given by $\mathcal{O} (K N)$.
This complexity is linear in the number $N$ of the fading blocks, and thus, it is much lower than the exponential complexity $\mathcal{O} ( M^{N} )$ of the exhaustive searching.

\setcounter{equation}{26}

So far, we have studied the problem (P2) under both the long-term and short-term power constraints.
Now, relaxing the short-term power constraint, we further investigate the problem (P2) only with the long-term power constraint to obtain the theoretically largest R-E region of the dynamic power splitting scheme.
Note that for the case of CSI, the theoretically best performance of the SWIPT system in the fading channel can be achieved when only the long-term power constraint is imposed \cite{Kang_3}.
Interestingly, in the following, we show that only with the long-term power constraint, it is possible to even further reduce the computational complexity to solve (P2) (while achieving the theoretically best performance).
\begin{lemma}
\label{lem_4}
The solution to (P2) only with the long-term power constraint is given by (\ref{P2_sol}) with $ P_{{\rm ID}, \nu}^{\texttt{NL}} $ and $P_{{\rm EH}, \nu}^{\texttt{NL}}$ replaced by $ P_{{\rm ID}, \nu}^{\texttt{NL}'} $ and $P_{{\rm EH}, \nu}^{\texttt{NL}'}$, respectively, where
\begin{subequations}
\label{P_NL_no_short}
\begin{align}
\label{P_ID_NL_no_short}
P_{{\rm ID}, \nu}^{\texttt{NL}'} &  =  \max \left\{  \frac{1}{\mu^{\texttt{NL}} } - \frac{\sigma^{2}}{h_{\nu}}, 0 \right\}, \\
\label{P_EH_NL_no_short}
P_{{\rm EH}, \nu}^{\texttt{NL}'}  & =  \begin{cases}  0 ,   &   \rm{for~Case~A.1'}  \\
    \begin{cases} P_{{\rm o}, \nu}^{A} ,  &  {\rm if} ~ \lambda^{\texttt{NL}} Q_{\nu}^{\texttt{NL}}  ( P_{{\rm o}, \nu}^{A} , 1 ) \\ & \quad >  \mu^{\texttt{NL}} P_{{\rm o}, \nu}^{A}  \\ 0, & {\rm otherwise}  \end{cases},  & \rm{for~Case~A.2'}  \\
    P_{{\rm o}, \nu}^{A}   ,   &   \rm{for~Case~A.3'}
\end{cases}.
\end{align}
\end{subequations}
In (\ref{P_EH_NL_no_short}), $ P_{{\rm o}, \nu}^{A} $ is given by (\ref{P_A_opt}).
Also, the Cases A.1$'$--A.3$'$ are given by
\begin{align}
\label{p_EH_NL_A_cases_no_short}
\begin{aligned}
& h_{\nu} \leq  x_{\rm{EH}, 1}^{\texttt{NL}'} && \quad {\rm for ~ Case ~ A.1'}  \\
& x_{\rm{EH}, 1}^{\texttt{NL}'} < h_{\nu} <  x_{\rm{EH}, 2}^{\texttt{NL}'} && \quad {\rm for ~ Case ~ A.2'} \\
& h_{\nu} \geq x_{\rm{EH}, 2}^{\texttt{NL}'} && \quad {\rm for ~ Case ~ A.3'} 
\end{aligned}
\end{align}
where
$x_{\rm{EH}, 1}^{\texttt{NL}'} = \frac{ 4 \mu^{\texttt{NL}} (1 - \Omega) }{ \lambda^{\texttt{NL}} P_{s} T a }$ and $x_{\rm{EH}, 2}^{\texttt{NL}'} = \frac{ \mu^{\texttt{NL}}  }{ \lambda^{\texttt{NL}} \Omega P_{s} T a  }$ $( > x_{\rm{EH}, 1}^{\texttt{NL}'} )$ are the roots of the equations $\mu^{\texttt{NL}} Z ( x ) = \frac{\lambda^{\texttt{NL}}}{4}$ and $\mu^{\texttt{NL}} Z ( x ) = \lambda^{\texttt{NL}} F_{\rm low}(x)$ with $P_{\rm low} = 0$, respectively. The constants $ \lambda^{\texttt{NL}} > 0 $ and $ \mu^{\texttt{NL}} > 0 $ are determined such that $\mathbb{E} \left[  Q_{\nu}^{\texttt{NL}} (  P_{\nu}^{\texttt{NL}} , \rho_{\nu}^{\texttt{NL}}  )  \right] = Q$ and $\mathbb{E} \left[  P_{\nu}^{\texttt{NL}}  \right]  = P_{\rm avg}$.
\end{lemma}
\begin{IEEEproof}
To relax the short-term power constraint, we take $P_{\max} \rightarrow \infty$.
From (\ref{P_ID_NL}), it follows that $ P_{{\rm ID}, \nu}^{\texttt{NL}} \rightarrow \max \left\{  \frac{1}{\mu^{\texttt{NL}} } - \frac{\sigma^{2}}{h_{\nu}}, 0 \right\} $ as $P_{\max} \rightarrow \infty$.
Also, from (\ref{P_EH_NL}), we have $P_{{\rm EH}, \nu}^{\texttt{NL}} \rightarrow \left[  P_{{\rm EH}, \nu}^{A} \right]_{0}^{ \infty }$ as $P_{\max} \rightarrow \infty$.
Since $ F_{\rm up} ( h_{\nu}  ) \rightarrow 0 $ as $P_{\rm up} \rightarrow \infty$,
the Cases A.3 and A.5 disappear from (\ref{p_EH_NL_A_cases}).
Also, the Cases A.1, A.2, and A.4 become the Cases A.1$'$, A.2$'$, and A.3$'$, respectively.
Thus, the result of (\ref{P_NL_no_short}) follows.
\end{IEEEproof}

From Lemma \ref{lem_4}, only with the long-term power constraint, the optimal solution to (P2) can be computed very efficiently.
Specifically, only the two positive scalars $ \lambda^{\texttt{NL}}$ and $ \mu^{\texttt{NL}}$ need to be determined numerically.
Consequently, the complexity is given by $\mathcal{O}(N)$, which is proportional only to $N$.

\subsection{Suboptimal Solution to (P2)}

The computational complexity of the optimal solution to (P2) in Theorem \ref{thm_3} is low; but, the complexity is proportional to both $N$ and $K$.
Thus, the complexity might not be low enough for certain practical applications.
On the other hand, the complexity of the solution in Lemma \ref{lem_4} is always low enough since its complexity is proportional only to $N$.
However, this solution can be used only when the short-term power constraint is relaxed (i.e., only the long-term power constraint is imposed).
In practice, it is often meaningful to consider the short-term power constraint because the short-term power is usually limited by the regulation bodies such as the FCC.
Therefore, in general, one needs to use the solution in Theorem \ref{thm_3}.
In this subsection, to overcome the above limitations, we present a suboptimal solution to (P2) in closed form without relaxing the short-term power constraint.
This solution turns out to be asymptotically optimal in the low SNR regime.

In order to derive a suboptimal solution, we define $ P_{{\rm EH}, \nu} = \rho_{\nu} P_{\nu} $, $\forall \nu$,
each of which denotes the power allocation for EH at each fading state.
Using the result of Theorem \ref{thm_3}, it can be shown that when $ h_{\nu} \leq  \mu^{\texttt{NL}} \sigma^{2} $ and $ P_{{\rm th}, \nu} \leq P_{{\rm EH}, \nu} \leq P_{\max} $, the optimization of (P2) reduces to the following optimization (the detailed proof is given in Appendix \ref{proof_thm_3}):
\begin{align}
\label{P_B_opt}
\underset{ P_{{\rm th}, \nu} < P_{{\rm EH}, \nu} \leq P_{\max} }{\texttt{max}} & \quad R_{\nu} ( P_{\max} - P_{{\rm EH}, \nu}, 0 )  + \lambda^{\texttt{NL}} Q_{\nu}^{\texttt{NL}} ( P_{{\rm EH}, \nu}, 1 ).
\end{align}
To carry out the above optimization, the value of $ P_{{\rm o}, \nu}^{ B} $ needs to be computed numerically by solving the nonlinear equation of (\ref{P_B_opt}),
which in turn increases the complexity.
To address the complexity issue and to obtain a closed form solution, we take the following approach:
we use the lower bound of the objective function of (\ref{P_B_opt}) for the optimization instead of the actual objective function.
To this end, in the following, we derive the lower bound of the objective function of (\ref{P_B_opt}).
\begin{lemma}
\label{lem_5}
The objective function of (\ref{P_B_opt}) is lower bounded by
\begin{align}
\label{obj_lb}
& R_{\nu} ( P_{\max} - P_{{\rm EH}, \nu}, 0 )  + \lambda^{\texttt{NL}} Q_{\nu}^{\texttt{NL}} ( P_{{\rm EH}, \nu}, 1 ) \nonumber\\
& \geq \left( 1 - \frac{ P_{{\rm EH}, \nu} }{ P_{\max} } \right) R_{\nu} ( P_{\max}, 0 )   + \lambda^{\texttt{NL}} Q_{\nu}^{\texttt{NL}} ( P_{{\rm EH}, \nu}, 1 ).
\end{align}
\end{lemma}
\begin{IEEEproof}
It follows that $ R_{\nu} ( P_{\max} - P_{{\rm EH}, \nu}, 0 ) \geq  \left( 1 - \frac{ P_{{\rm EH}, \nu} }{ P_{\max} } \right) R_{\nu} ( P_{\max}, 0 )  $
since $ R_{\nu} ( P_{\max} - P_{{\rm EH}, \nu}, 0 ) = \log_{2} \left(  1 + \frac{h_{\nu} ( P_{\max} - P_{{\rm EH}, \nu} ) }{\sigma^{2}} \right) = \log_{2} \left(  1 + \frac{h_{\nu} P_{\max}  }{\sigma^{2}} \left( 1 - \frac{ P_{{\rm EH}, \nu} }{ P_{\max} } \right) \right) $ is concave in $ \left( 1 - \frac{ P_{{\rm EH}, \nu} }{ P_{\max} } \right) $ \cite{Boyd}.
\end{IEEEproof}

The lower bound of (\ref{obj_lb}) in Lemma \ref{lem_5} becomes tight when $ P_{{\rm EH}, \nu} = P_{\max} $.
Replacing the objective function of (\ref{P_B_opt}) with the lower bound of (\ref{obj_lb}),
we have the following optimization problem:
\begin{align}
{\rm (P2'):}  \quad  \underset{ P_{{\rm th}, \nu} < P_{{\rm EH}, \nu} \leq P_{\max} }{\texttt{max}} & \quad  \bigg\{ \left( 1 - \frac{ P_{{\rm EH}, \nu} }{ P_{\max} } \right) R_{\nu} ( P_{\max}, 0 ) \nonumber\\
 & \quad + \lambda^{\texttt{NL}} Q_{\nu}^{\texttt{NL}} ( P_{{\rm EH}, \nu}, 1 )  \bigg\}.
\end{align}

\begin{figure*}[t]
\begin{align}
\label{p_EH_NL_B_so}
& \big[ P_{{\rm EH}, \nu}^{B'} \big]_{P_{\rm low}}^{ P_{\rm up} }  \nonumber\\
&   = \begin{cases}  P_{\rm low}  ,   &   {\rm for~Case~B . 1'}   \\
    \begin{cases}  P_{{\rm so}, \nu}^{ B}  ,  &  {\rm if} ~  R_{\nu} ( P_{\rm up} - P_{{\rm so}, \nu}^{ B} , 0 ) + \lambda^{\texttt{NL}} Q_{\nu}^{\texttt{NL}} ( P_{{\rm so}, \nu}^{ B} , 1 )  > R_{\nu} ( P_{\rm up} - P_{\rm low} , 0 ) + \lambda^{\texttt{NL}} Q_{\nu}^{\texttt{NL}} ( P_{\rm low} , 1 ) \\ P_{\rm low}, & {\rm otherwise}  \end{cases},  &  {\rm for~Case~B . 2'} \\
    \begin{cases}  P_{\rm up} ,  &  {\rm if}  ~  \lambda^{\texttt{NL}} Q_{\nu}^{\texttt{NL}} ( P_{\rm up} , 1 ) > R_{\nu} ( P_{\rm up} - P_{\rm low} , 0 ) + \lambda^{\texttt{NL}} Q_{\nu}^{\texttt{NL}} ( P_{\rm low} , 1 )  \\ P_{\rm low}, & {\rm otherwise}  \end{cases},  & {\rm for~Case~B . 3'} \\
    P_{{\rm so}, \nu}^{ B}   ,   &   {\rm for~Case~B . 4'}  \\
    P_{\rm up}   ,   &   {\rm for~Case~B . 5'}
\end{cases}.
\end{align}
\hrulefill
\begin{align}
\label{p_EH_NL_B_so_cases}
&  Z' ( h_{\nu} ) \geq  \tfrac{\lambda^{\texttt{NL}}}{4}    && \quad {\rm for ~ Case ~ B . 1'}   \nonumber\\
& \max \left\{ \lambda^{\texttt{NL}} F ( h_{\nu} , P_{\rm up} ), \lambda^{\texttt{NL}} F ( h_{\nu} , P_{\rm low} )  \right\} <  Z' ( h_{\nu} )  < \tfrac{\lambda^{\texttt{NL}}}{4}  && \quad {\rm for ~ Case ~ B . 2'}  \nonumber\\
& \lambda^{\texttt{NL}} F ( h_{\nu} , P_{\rm low} )  \leq   Z' ( h_{\nu} )  \leq \lambda^{\texttt{NL}} F ( h_{\nu} , P_{\rm up} )  && \quad {\rm for ~ Case ~ B . 3'}  \nonumber\\
& \lambda^{\texttt{NL}} F ( h_{\nu} , P_{\rm up} )  \leq   Z' ( h_{\nu} )  \leq \lambda^{\texttt{NL}} F ( h_{\nu} , P_{\rm low} )  && \quad {\rm for ~ Case ~ B . 4'}  \nonumber\\
&  Z' ( h_{\nu} )  \leq   \min \left\{ \lambda^{\texttt{NL}} F ( h_{\nu} , P_{\rm up} ), \lambda^{\texttt{NL}} F ( h_{\nu} , P_{\rm low} )  \right\}  && \quad {\rm for ~ Case ~ B . 5'}.
\end{align}
\hrulefill
\end{figure*}

In the following, the solution to (P2$'$) is derived.
\begin{theorem}
\label{thm_4}
The solution to (P2$'$) is given by $ \big[ P_{{\rm EH}, \nu}^{B'} \big]_{  P_{{\rm th}, \nu} }^{ P_{\max} } $,
where $ \big[ P_{{\rm EH}, \nu}^{B'} \big]_{P_{\rm low}}^{ P_{\rm up} } $ is given by (\ref{p_EH_NL_B_so}) (shown at the top of the next page).
In (\ref{p_EH_NL_B_so}), the Cases B.1$'$--B.5$'$ are given by (\ref{p_EH_NL_B_so_cases}) (shown at the top of the next page),
where $ Z' (x) =   \frac{ 1 - \Omega  }{ P_{s} T a x P_{\max} } \log_{2} \left(  1 + \frac{x P_{\max}}{\sigma^{2}} \right)$.
Also, $P_{{\rm so}, \nu}^{ B}$ is given by
\begin{align}
\label{P_B_so}
P_{{\rm so}, \nu}^{ B}  =  \frac{1}{h_{\nu}} \left( - \frac{1}{a}  \ln \left(  \frac{ 2 }{  1 + \sqrt{ 1 - \frac{ 4 Z' ( h_{\nu} ) }{ \lambda^{\texttt{NL}} } }  } - 1 \right) + b \right).
\end{align}
\end{theorem}
\begin{IEEEproof}
The result can be proved by following the similar procedures in Appendix \ref{proof_thm_3}, where $P_{{\rm so}, \nu}^{ B} $ in (\ref{P_B_so}) is given by the solution of the equation $ \Psi_{\nu} ( x, 1  ) =  \frac{1}{2} + \sqrt{ \frac{1}{4} - \frac{Z'(h_{\nu})}{\lambda^{\texttt{NL}}} } $ over $ x \in \Big( \frac{b}{h_{\nu}},  P_{\rm up}  \Big) $.
\end{IEEEproof}

Using Theorem \ref{thm_4}, the suboptimal solution to (P2) can be obtained from the result of Theorem \ref{thm_3} with $ \big[ P_{{\rm EH}, \nu}^{B} \big]_{  P_{{\rm th}, \nu} }^{ P_{\max} } $ replaced by $ \big[ P_{{\rm EH}, \nu}^{B'} \big]_{  P_{{\rm th}, \nu} }^{ P_{\max} } $.
Note that the computational complexity of this solution is very low: only the two positive scalars $ \lambda^{\texttt{NL}}$ and $ \mu^{\texttt{NL}}$ need to be computed numerically.
Thus, the complexity is given by $\mathcal{O} (N)$, which is proportional only to $N$ and is even lower than the complexity of the optimal solution in Theorem \ref{thm_3}.
However, the performance of the solution using Theorem \ref{thm_4} is generally suboptimal since the lower bound of (\ref{obj_lb}) is used rather than the exact value.
Fortunately and interestingly, this solution achieves the asymptotic optimality in the low SNR regime, which is presented in the following.
\begin{lemma}
\label{lem_optimality_2}
When $ \frac{ P_{\max} }{ \sigma^{2} } \rightarrow 0 $,
the solution in Theorem \ref{thm_3} with $ \big[ P_{{\rm EH}, \nu}^{B} \big]_{  P_{{\rm th}, \nu} }^{ P_{\max} } $ replaced by $ \big[ P_{{\rm EH}, \nu}^{B'} \big]_{  P_{{\rm th}, \nu} }^{ P_{\max} } $ is optimal to (P2).
\end{lemma}
\begin{IEEEproof}
For the proof, it is sufficient to show that the gap between the actual objective value of (\ref{P_B_opt}) and the lower bound of (\ref{obj_lb}) goes to zero as $ \frac{ P_{\max} }{ \sigma^{2} } \rightarrow 0 $.
When $ \frac{ P_{\max} }{ \sigma^{2} } \rightarrow 0 $, it follows that $ R_{\nu} ( P_{\max} - P_{{\rm EH}, \nu}, 0 )  \rightarrow  \frac{ h_{\nu} (  P_{\max} - P_{{\rm EH}, \nu} ) }{ \sigma^{2} \ln 2 }  $
and $ \left( 1 - \frac{ P_{{\rm EH}, \nu} }{ P_{\max} } \right) R_{\nu} ( P_{\max}, 0 ) \rightarrow \frac{ h_{\nu} (  P_{\max} - P_{{\rm EH}, \nu} ) }{ \sigma^{2} \ln 2 }  $ \cite{Tse}.
Thus, $ R_{\nu} ( P_{\max} - P_{{\rm EH}, \nu}, 0 ) - \left( 1 - \frac{ P_{{\rm EH}, \nu} }{ P_{\max} } \right) R_{\nu} ( P_{\max}, 0 ) \rightarrow 0$ as $ \frac{ P_{\max} }{ \sigma^{2} } \rightarrow 0 $.
\end{IEEEproof}

Although we can mathematically show the optimality of the solution in Theorem \ref{thm_4} only in the low SNR region,
the numerical results in the next section will demonstrate that the performance of this solution is very close to the optimal performance even in the moderate to high SNR range.

\subsection{Comparisons to Linear EH}

In this subsection, to obtain further insights, the proposed scheme is compared to the existing dynamic power splitting scheme for the linear EH \cite[Proposition 4.2]{Liu}:
\begin{align}
\label{rho_P_sol_L}
( P_{\nu}^{\texttt{L}} , \rho_{\nu}^{\texttt{L}} )& =  \begin{cases}  \begin{cases} ( P_{\max} , \rho_{{\rm o}, \nu}^{\texttt{L}} )  ,  &  {\rm if}  ~  h_{\nu}  >  x_{1}^{\texttt{L}}  \\ \left( \left[ P_{{\rm o}, \nu}^{\texttt{L}} \right]_{0}^{P_{\max}} , 0 \right), & {\rm otherwise}  \end{cases}, & {\rm if~}  x_{2}^{\texttt{L}} \leq x_{1}^{\texttt{L}} \\ \begin{cases} ( P_{\max} , \rho_{{\rm o}, \nu}^{\texttt{L}} )  ,  &  {\rm if}  ~  h_{\nu}  >  x_{2}^{\texttt{L}}  \\ \left( \left[ P_{{\rm o}, \nu}^{\texttt{L}} \right]_{0}^{\infty} , 0 \right) , & {\rm otherwise}  \end{cases}, & {\rm if~}  x_{2}^{\texttt{L}}  < x_{1}^{\texttt{L}} \end{cases}
\end{align}
where
$ \left[ P_{{\rm o}, \nu}^{\texttt{L}} \right]_{P_{\rm low}}^{P_{\rm up}}  = \min \left\{ \max \left\{ \frac{1}{\mu^{\texttt{L}}} - \frac{\sigma^{2}}{h_{\nu}} , P_{\rm low} \right\} , P_{\rm up} \right\} $
and $\rho_{{\rm o}, \nu}^{\texttt{L}}  = 1 - \frac{1}{h_{\nu} P_{\max}} \left( \frac{1}{ \zeta \lambda^{\texttt{L}} } - \sigma^{2} \right)$.
Also, $x_{1}^{\texttt{L}} = \frac{1}{ P_{\max} }  \left( \frac{1}{ \zeta \lambda^{\texttt{L}} } - \sigma^{2} \right)$ and $x_{2}^{\texttt{L}} = \frac{\mu^{\texttt{L}}}{\zeta \lambda^{\texttt{L}}}$.
The constants $ 0 < \lambda^{\texttt{L}} \leq \frac{1}{\zeta \sigma^{2}} $ and $\mu^{\texttt{L}} > 0$ are determined such that $\mathbb{E} \left[  Q_{\nu}^{\texttt{L}} (  P_{\nu}^{\texttt{L}} , \rho_{\nu}^{\texttt{L}}  )  \right] = Q$ and $\mathbb{E} \left[  P_{\nu}^{\texttt{L}}  \right]  = P_{\rm avg}$.

Comparing the proposed and existing schemes and considering the practical distance-based channel model of \cite{Huang15} discussed in Section \ref{sec_P_EH_comp}, we gain the further useful insights related to the distance as follows:
\begin{itemize}
\item From the results of (\ref{P2_sol}), (\ref{P_NL_no_short}), and (\ref{rho_P_sol_L}), we can make the following observations.
    In the proposed and existing schemes, the transmission is turned off when the distance $d$ is very large (or the channel gain $h$ is very small) to save the transmit power.
    However, in the small to moderate range of the distances (or the moderate to large range of the channel gains), the proposed and existing schemes become very different.
    For example, in the optimal dynamic power splitting scheme for the linear EH, the transmit power is allocated to maximize the amount of information transfer when the distance $d$ (or the channel gain $h$) is \textit{moderate}.    
    On the other hand, in the optimal dynamic power splitting scheme for the nonlinear EH, the transmit power is allocated to maximize the amount of information transfer when the distance $d$ (or the channel gain $h$) is \textit{small} or \textit{large}.
\end{itemize}

\subsection{Extension to the Case of Harvested Energy Maximization}
Throughout this paper, we studied the rate maximization with the harvested energy constraint,
which is important and useful for the SWIPT applications with fixed energy consumption at the receiver.
On the other hand, for the fixed-rate SWIPT applications, it is important and useful to maximize the harvested energy with the rate constraint.
In this subsection, we extend our analysis to the case of harvested energy maximization.
To this end, without loss of generality, we consider the problem of maximizing the average harvested energy in the case of CSI with the average rate constraint
as follows: $\underset{ \subalign{ & 0 \leq \rho_{\nu} \leq 1,  \forall \nu, \\  & 0 \leq P_{\nu} \leq P_{\max} , \forall \nu  } }{\texttt{max}} ~ \mathbb{E} \left[  Q_{\nu}^{\texttt{NL}} (  P_{\nu} , \rho_{\nu}  )  \right] ~
\texttt{s.t.}  ~ \mathbb{E} \left[  R_{\nu} ( P_{\nu} , \rho_{\nu}  )  \right] \geq R, ~  \mathbb{E} \left[  P_{\nu}  \right]  \leq P_{\rm avg}$,
where $R$ denotes the threshold for the average rate.
This problem can be solved by extending our analytical approaches to solve the problem (P2).
Specifically, following the similar procedures in Appendix \ref{proof_lems_1_2}, 
it can be shown that the time-sharing condition holds. 
Thus, using the Lagrange duality method and following the similar procedures in Appendix \ref{proof_thm_3},
the solution can be obtained from the results in Theorem \ref{thm_3} with $\lambda^{\texttt{NL}}$ and $\mu^{\texttt{NL}}$ replaced by $\frac{1}{\widetilde{\lambda}^{\texttt{NL}}}$ and $\frac{\widetilde{\mu}^{\texttt{NL}}}{\widetilde{\lambda}^{\texttt{NL}}} $, respectively,
where the dual variables $\widetilde{\lambda}^{\texttt{NL}} > 0$ and $\widetilde{\mu}^{\texttt{NL}} > 0$ are determined such that 
the average rate constraint and the average transmit power constraint are all satisfied with equality.

\begin{figure*}[!t]
\centering
{
    \includegraphics[width=0.7\textwidth]{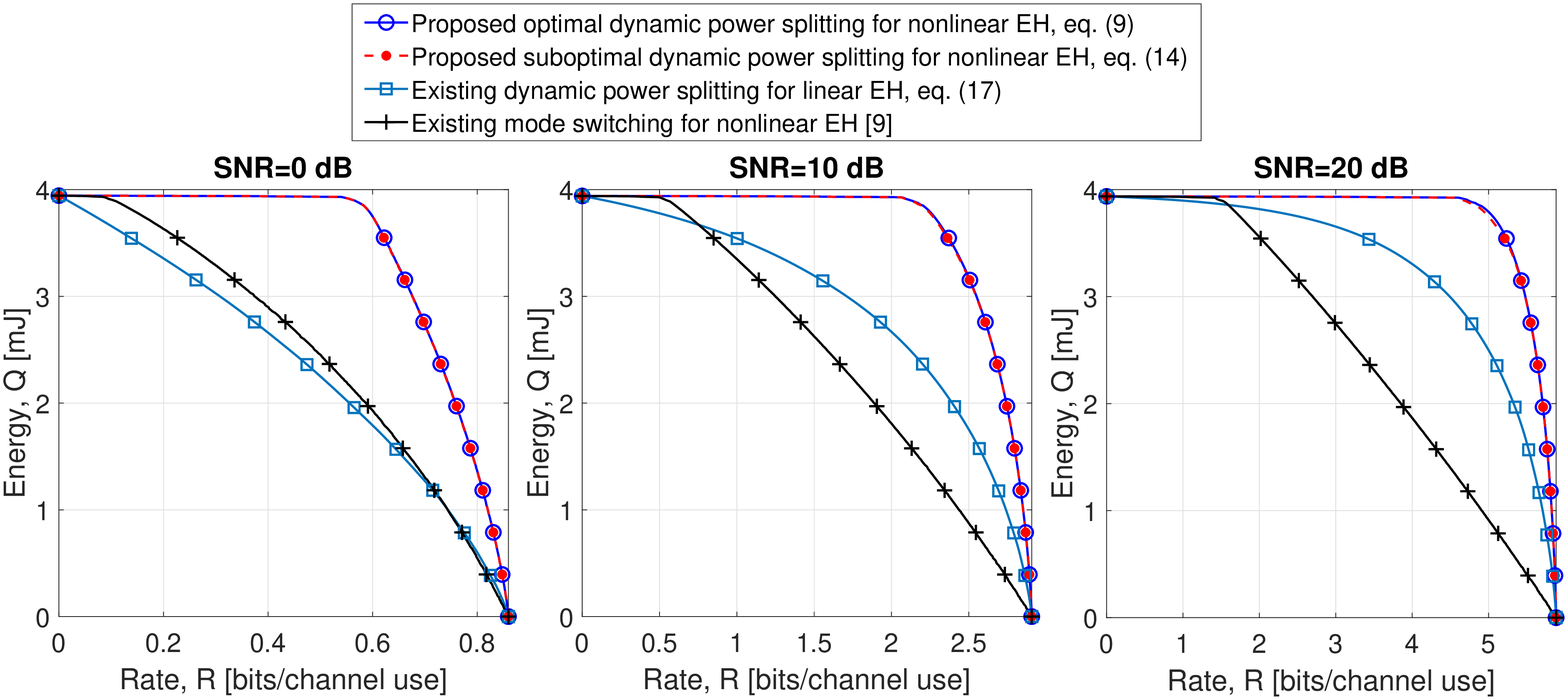}
    \caption{
    R-E regions of the proposed and existing schemes for the case of CSIR.
    }
    \label{fig2}
}
\end{figure*}

\begin{figure*}[!t]
\centering
{
    \includegraphics[width=0.7\textwidth]{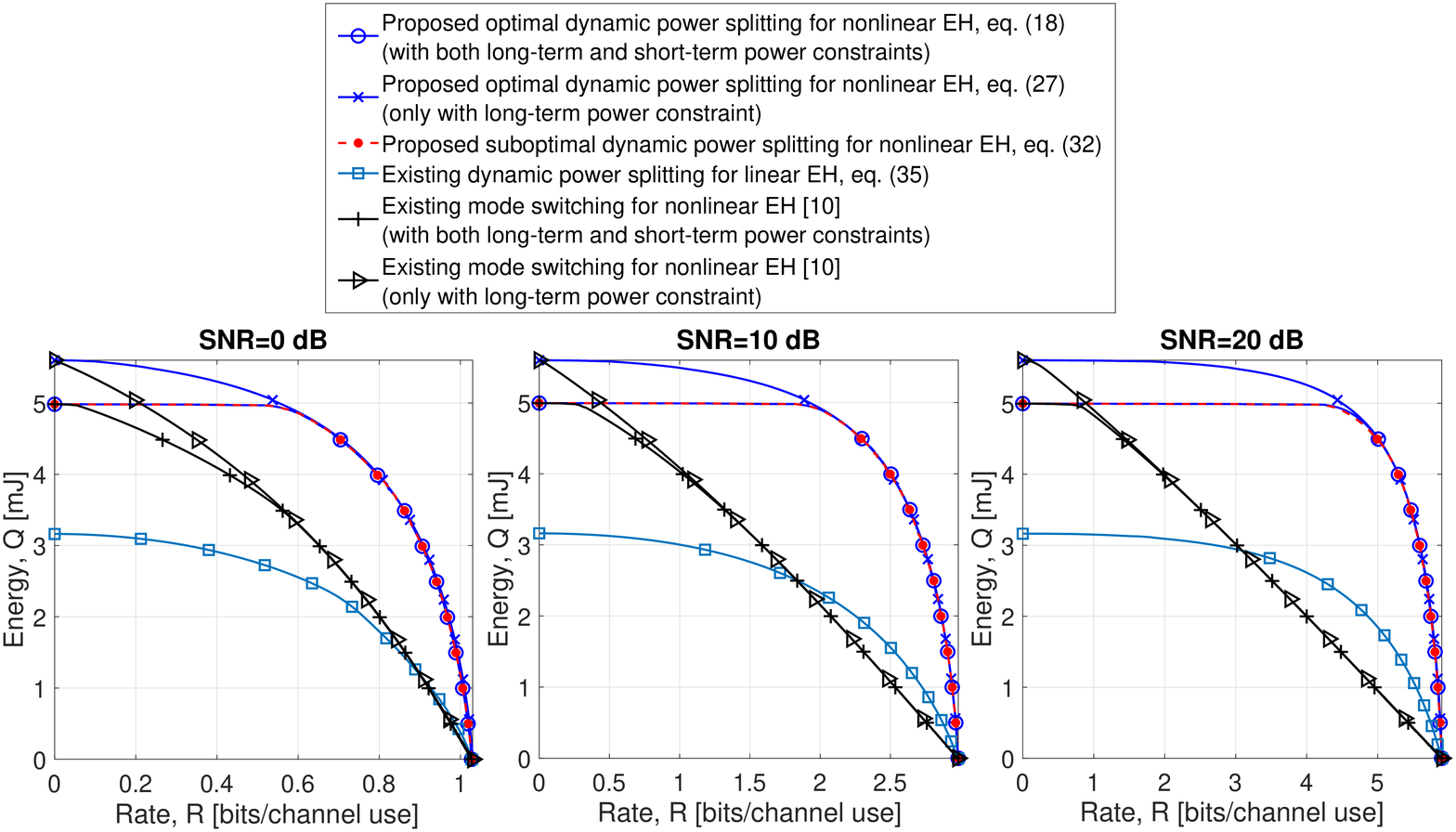}
    \caption{
    R-E regions of the proposed and existing schemes for the case of CSI.
    }
    \label{fig3}
}
\end{figure*}

\begin{figure*}
    \centering
    \subfigure[The CSIR case]
    {
        \includegraphics[width=0.4\textwidth]{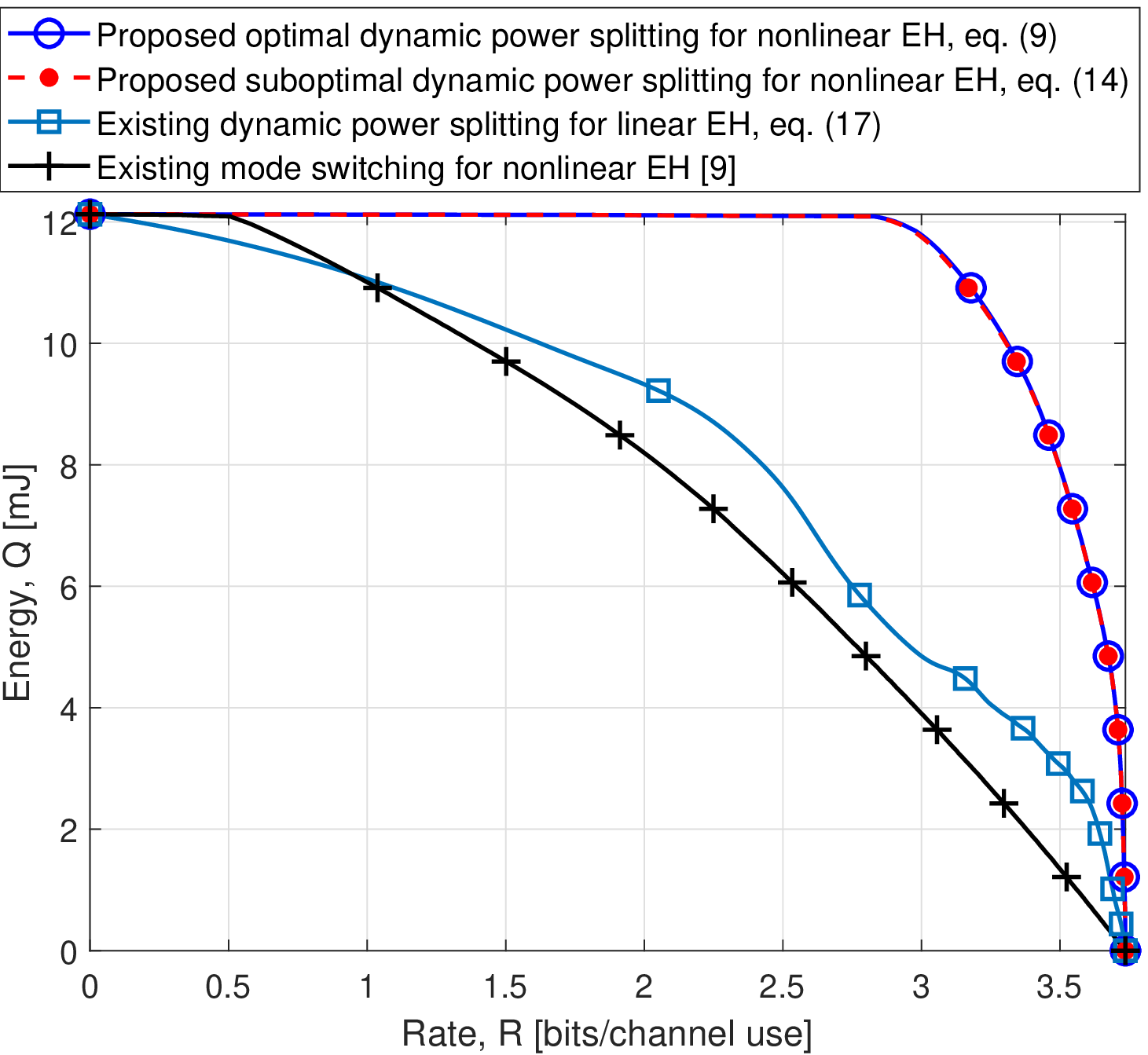}
        \label{fig4_a}
    }
    \subfigure[The CSI case]
    {
        \includegraphics[width=0.4\textwidth]{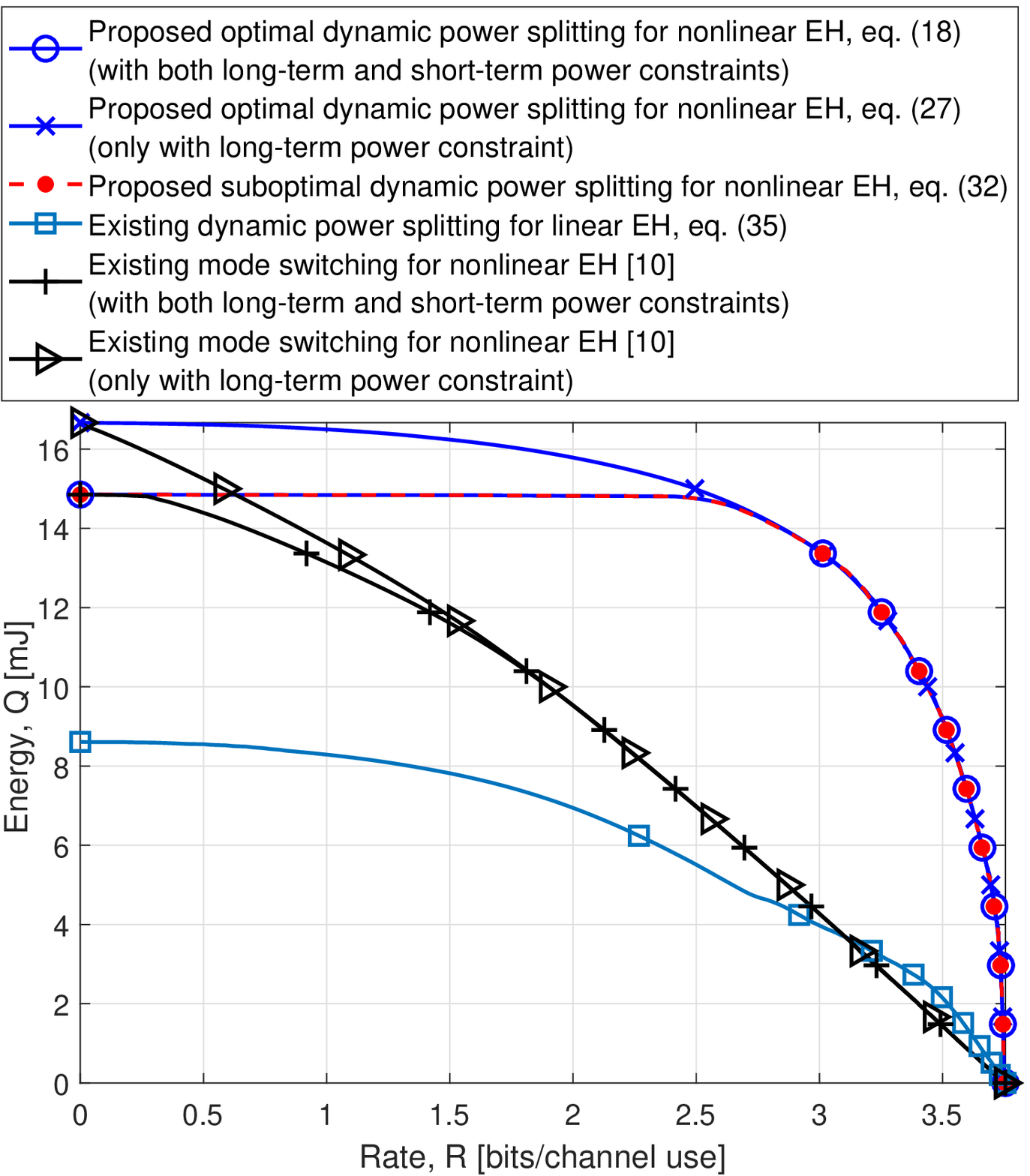}
        \label{fig4_b}
    }
    \caption{R-E regions of the proposed and existing schemes with the receiver mobility when $ \rm{SNR} = 15 $ dB.}
    \label{fig4}
\end{figure*}

\section{Numerical Results}
\label{numerical_result}

In this section, the performance of the proposed scheme is demonstrated through the numerical results and it is compared to that of the existing scheme of \cite{Liu}.
We also present the performance of the existing mode switching schemes developed in \cite{Kang_2} and \cite{Kang_3} for the nonlinear EH with CSIR and CSI, respectively.
We evaluate the R-E tradeoff performance of all the compared schemes using the practical nonlinear EH model of (\ref{Q_NL});
that is, the R-E regions of the schemes are defined as (\ref{R_E_region}) or (\ref{R_E_region_2}).
Unless stated otherwise, the following setting is used in the simulations.
The block duration is normalized to unity, i.e., $T = 1$ second.
Also, we set $a = 6400$, $b = 0.003$,
$P = P_{\rm avg} = 2 $ W, $P_{\max} = 4$ W, and $ P_{s} = \mathbb{E} [ h_{\nu} ] P_{\rm avg} $.
The average (receive) SNR is defined as ${\rm SNR} = \frac{ \mathbb{E} [ h_{\nu} ] P_{\rm avg} } {\sigma^{2}}$.
We generate $N = 10^{5}$ independent fading blocks according to the Rician channel model: $\eta_{\nu} = \sqrt{\frac{\alpha}{\alpha+1}} \bar{\eta}  + \sqrt{\frac{1}{\alpha+1}} \tilde{\eta}_{\nu} $, $\forall \nu$, where $\eta_{\nu}$ is the channel at the fading state $\nu$. Thus, the channel power gain $h_{\nu}$ at that state is given by $h_{\nu} = | \eta_{\nu} |^{2}$.
Also, $\alpha$ is the Rician factor, and $\bar{\eta}$ and $\tilde{\eta}_{\nu} \sim \mathcal{CN} (0, \sigma_{h}^{2})$ are the line-of-sight and scattering components, respectively.
We set $\alpha = 1$ and $| \bar{\eta} |^{2} = \sigma_{h}^{2} = - 28$ dBW.
In the existing dynamic power splitting scheme of \cite{Liu} for the linear EH, we set $\zeta = 1$.

In Figs. \ref{fig2} and \ref{fig3}, the R-E regions of the proposed and existing schemes are shown for the cases of CSIR and CSI, respectively, when $ {\rm SNR} \in \{ 0, 10, 20 \} $ dB. From Figs. \ref{fig2} and \ref{fig3}, one can see that the proposed schemes significantly outperform the existing schemes,
because the proposed scheme is optimal in the sense of maximizing the R-E region for the nonlinear EH.
In particular, since the existing dynamic power splitting scheme for the idealistic linear EH is strictly suboptimal for the case of practical nonlinear EH,
its R-E region is much smaller than the proposed scheme.
Also, the performance of the existing mode switching scheme of \cite{Kang_2} or \cite{Kang_3} is much worse than the proposed scheme 
because it can be considered as a simplified special case of the proposed scheme with $\rho_{\nu} \in \{ 0, 1 \}$, $\forall \nu$.
For the case of CSI, only with the long-term power constraint, the proposed optimal scheme yields the best performance.
As expected from Lemma \ref{lem_optimality}, the proposed suboptimal scheme provides the optimal performance in the low SNR range (e.g., when $ {\rm SNR} = 0 $ dB).
Interestingly, however, even in the moderate and high SNR ranges (e.g., when $ {\rm SNR} \in \{ 10, 20 \} $ dB),
this scheme performs very well and its performance is almost the same as the optimal performance.



Considering the practical applications such as the Internet-of-Things (IoT) with implanted bio and/or wearable sensors, in Fig. \ref{fig4}, we investigate the impact of the receiver mobility on the R-E tradeoff performance. 
In Fig. \ref{fig4}, the R-E regions of the proposed and existing schemes are shown for the cases of CSIR and CSI when $ {\rm SNR} = 15 $ dB.
In this figure, we use the practical distance-based channel model of \cite{Huang15} considered in Section \ref{sec_P_EH_comp}.
Assuming that the receiver is a small sensor, we set $a_{t} = 0.5$ m, $a_{r} = 0.01$ m, and $f_{c} = 2.4$ GHz \cite{Kim16}--\cite{Kang}.
Also, considering the mobility of the receiver, we set the distance $d_{\nu}$ in the current fading state $\nu$ as follows: $d_{\nu} = d_{\nu'} + \beta_{\nu} v_{\max} T$,
where $d_{\nu'}$ is the distance in the previous fading state $\nu'$; $v_{\max}$ the maximum speed; and $\beta_{\nu}$ the parameter accounting for the directionality.
We set the initial distance to $15$ m and we select $\beta_{\nu}$ randomly over the range $[-1,1]$. Also, we set $v_{\max} = 0.1$ m/s.
From Fig. \ref{fig4}, it can be seen that in the scenario of the receiver mobility, the proposed schemes perform much better than the existing schemes,
which clearly shows the significant benefit of the proposed schemes over the existing schemes in the practical applications.
Note that since the R-E region for the practical nonlinear EH model of (\ref{Q_NL}) is generally nonconvex as demonstrated in \cite{Kang}, \cite{Xiong},
the R-E region of the existing dynamic power splitting scheme for the linear EH has a nonconvex shape.



\begin{figure*}
    \centering
    \subfigure[For different $P_{s}$ when $a = 6400$ and $b = 0.003$]
    {
        \includegraphics[width=0.31\textwidth]{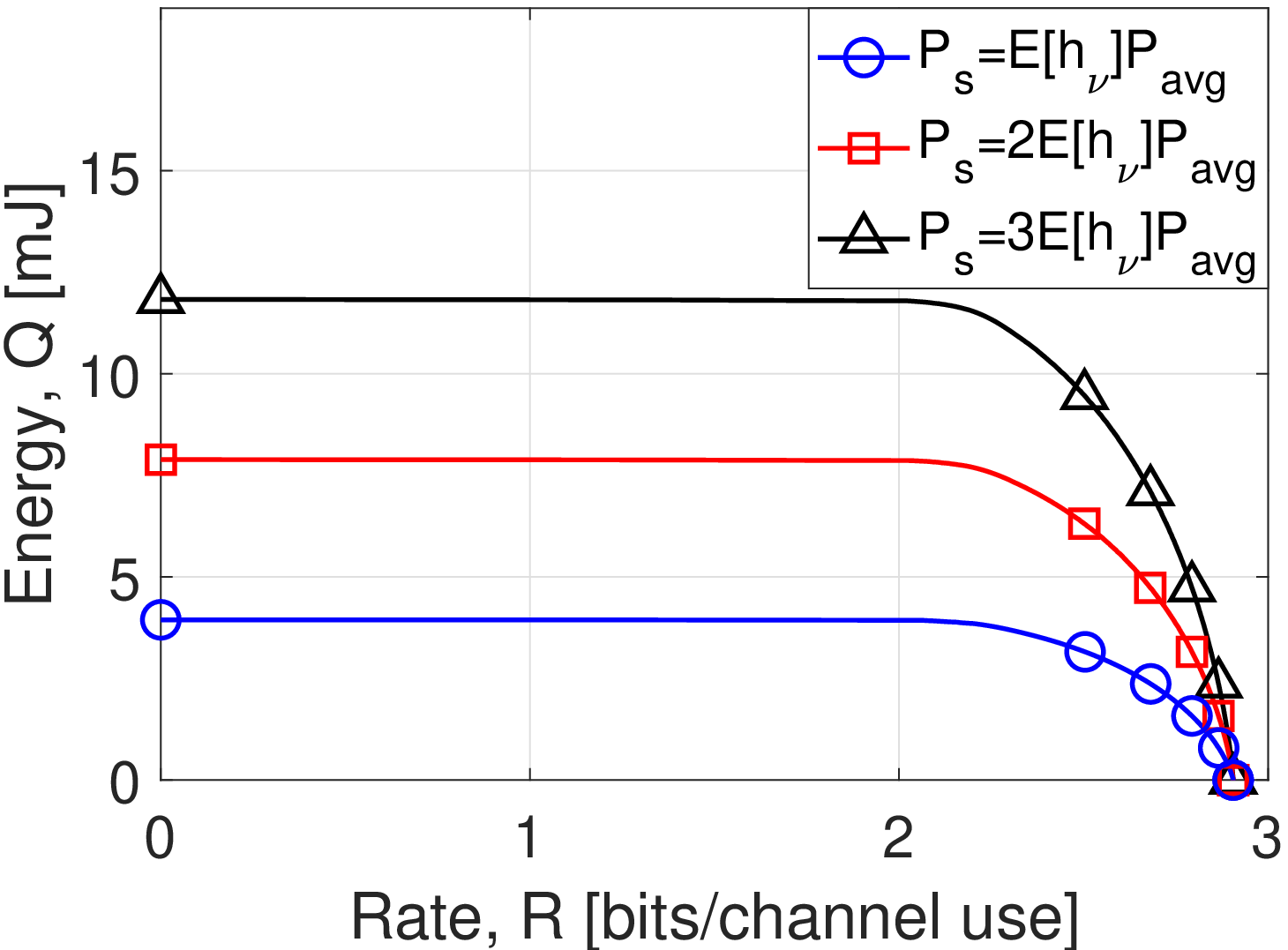}
        \label{fig5_a}
    }
    \subfigure[{For different $a$ when $P_{s} = \mathbb{E} [ h_{\nu} ] P_{\rm avg}$ and $b = 0.003$}]
    {
        \includegraphics[width=0.31\textwidth]{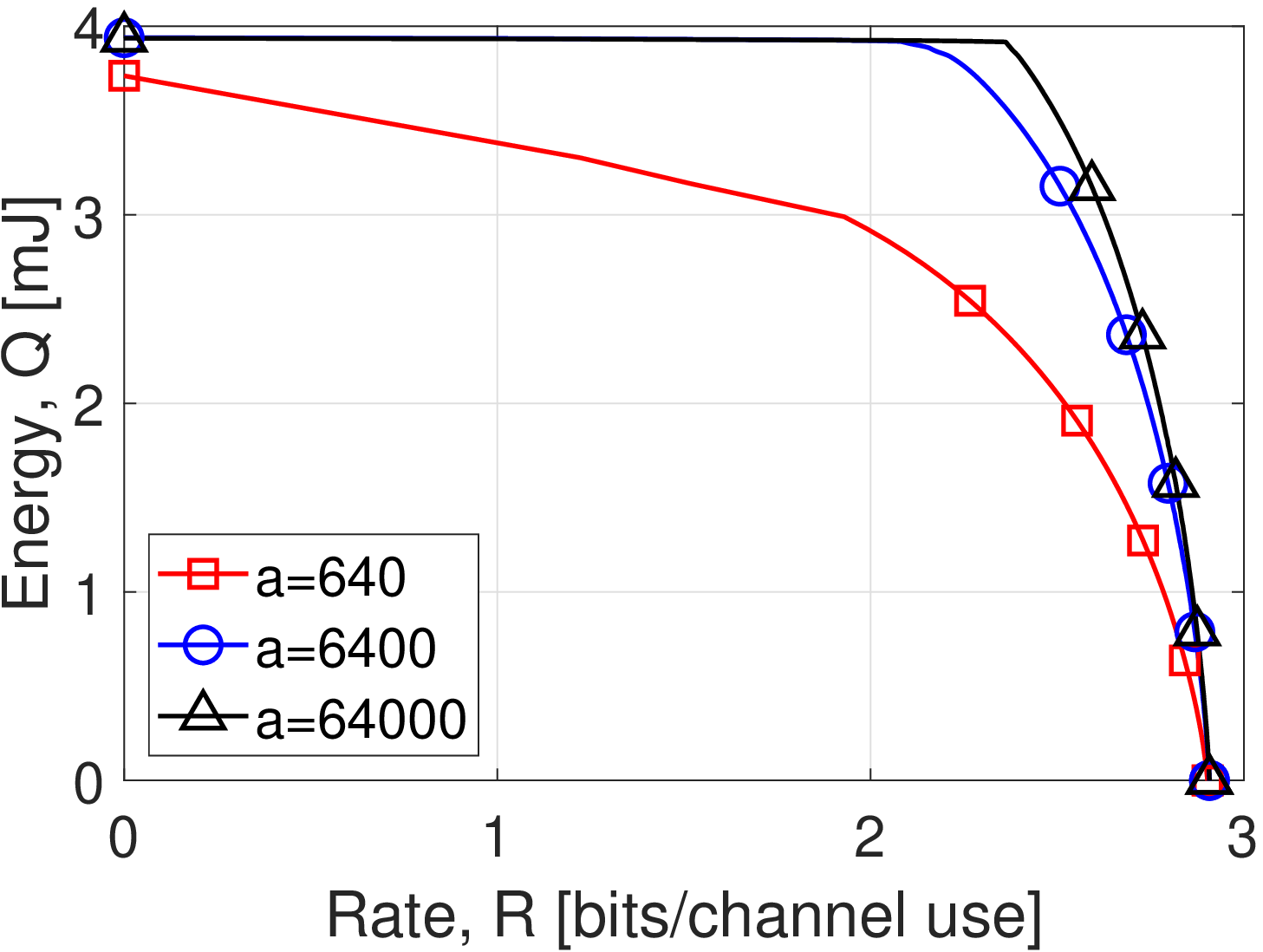}
        \label{fig5_b}
    }
    \subfigure[{For different $b$ when $P_{s} = \mathbb{E} [ h_{\nu} ] P_{\rm avg}$ and $a = 6400$}]
    {
        \includegraphics[width=0.31\textwidth]{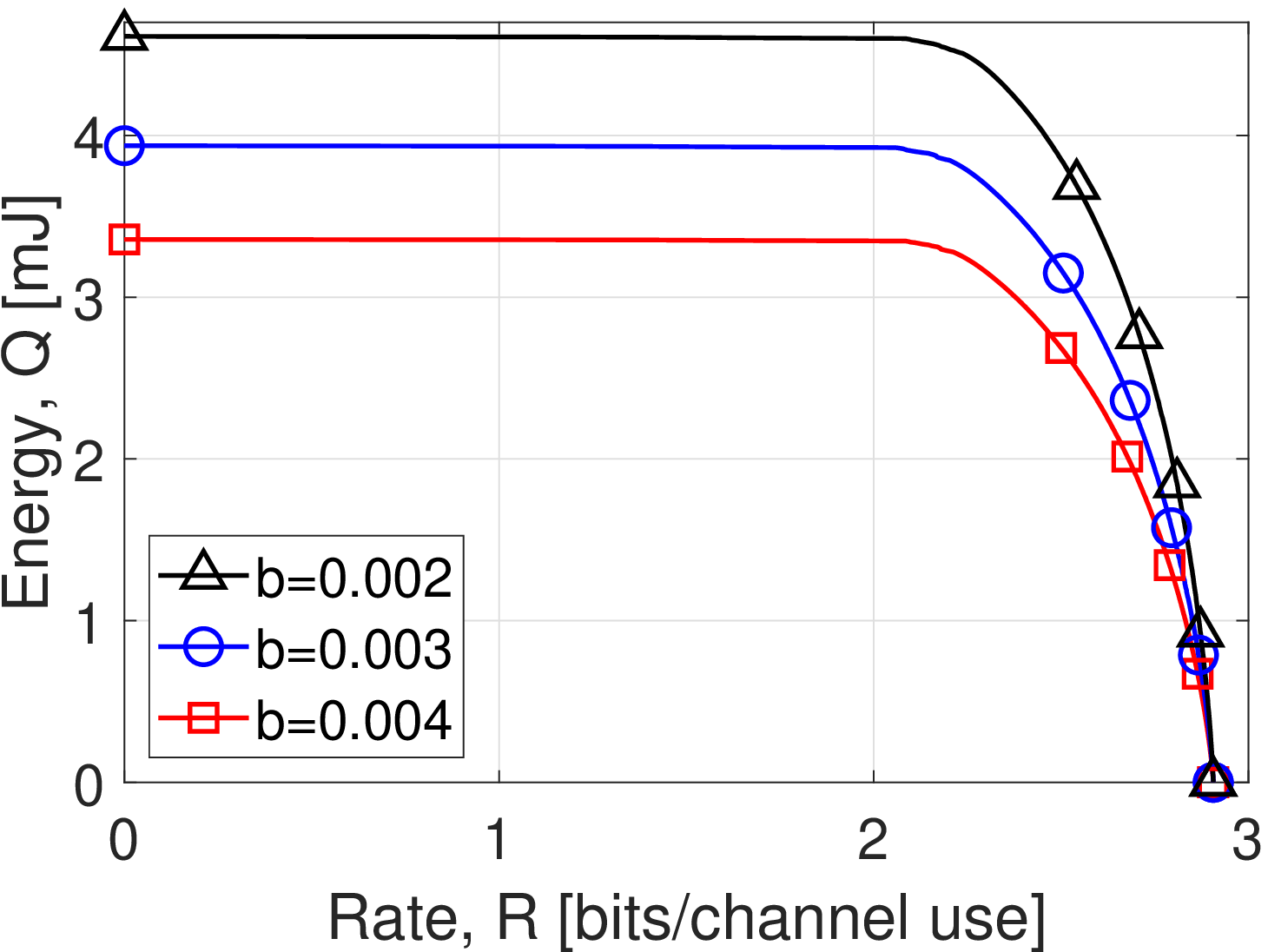}
        \label{fig5_c}
    }
    \caption{R-E regions of the proposed scheme in the case of CSIR for different values of $P_{s}$, $a$, and $b$.}
    \label{fig5}
\end{figure*}

In Fig. \ref{fig5}, we investigate the impacts of the parameters of the nonlinear EH model on the R-E tradeoff performance.
In this figure, in the case of the CSIR, we plot the R-E region of the proposed scheme for different values of $P_{s}$, $a$, and $b$.
The other parameters are the same as those in Fig. \ref{fig2}.
From Fig. \ref{fig5}, it can be seen that when $P_{s}$ increases (or decreases), $a$ increases (or decreases), and/or $b$ decreases (or increases),
the R-E tradeoff performance of our proposed scheme is improved (or degraded).

\section{Conclusion}

In this paper, we studied the dynamic power splitting for the SWIPT system with nonlinear EH in the ergodic fading channel in the sense of maximizing the R-E region.
We first developed the optimal and suboptimal dynamic power splitting scheme for the cases of CSIR,
and then, we developed the optimal and suboptimal schemes for case of CSI.
Comparing the proposed schemes to the existing schemes, the useful insights were obtained.
We also extended the analysis to the cases of the partial CSI at the transmitter and the harvested energy maximization.
The numerical results showed that the proposed schemes considerably outperformed the existing schemes
and the proposed suboptimal scheme performed very close to the optimal scheme.

\appendices

\numberwithin{equation}{section}




\section{Proofs of Lemmas \ref{lem_1} and \ref{lem_2}}
\label{proof_lems_1_2}

In this proof, we first prove the result of Lemma \ref{lem_2}. Then we prove the result of Lemma \ref{lem_1}.

\subsection{Proof of Lemma \ref{lem_2}}

Let $h_{k}$ denote the channel power gain at the $k$th block, $k=1,\cdots, N$.
Then, due to the law of large numbers, it follows that
$  \frac{1}{N} \sum_{k=1}^{N}  R_{k} ( P_{k} , \rho_{k} ) \rightarrow \mathbb{E} [  R_{\nu} (  P_{\nu } , \rho_{\nu } )  ] $,
$  \frac{1}{N} \sum_{k=1}^{N}  Q_{k}^{\texttt{NL}} ( P_{k} , \rho_{k} ) \rightarrow \mathbb{E} [  Q_{\nu}^{\texttt{NL}} (  P_{\nu } , \rho_{\nu } )  ] $,
and $  \frac{1}{N} \sum_{k=1}^{N} P_{k} \rightarrow \mathbb{E} [  P_{ \nu}  ] $ as $N \rightarrow \infty$,
where $R_{k} (\cdot) $ and $Q_{k}^{\texttt{NL}} (\cdot) $ are given by (\ref{R}) and (\ref{Q_NL}), respectively, with $h_{\nu}$ replaced by $h_{k}$.
Let $ \{ P_{x, k}^{*}, \rho_{ x, k}^{*} \}$ and $\{ P_{y, k}^{*}, \rho_{y, k}^{*} \}$ denote the solutions to (P2) with $ (Q, P_{\rm avg} ) =  ( Q_{x} , \overline{P}_{x} ) $ and $(Q, P_{\rm avg} )= (Q_{y} , \overline{P}_{y} )$, respectively.
Then, for (P2), a feasible point $ \{ P_{z, k} , \rho_{z, k} \} $ to satisfy the time-sharing condition always exists as follows:
\begin{align}
\label{A1}
( P_{z,k}, \rho_{z,k} ) = \begin{cases} ( P_{x,k}^{*},  \rho_{x,k}^{*} ) ,  &   k=1,\cdots, \theta N \\  ( P_{y,k}^{*},  \rho_{y,k}^{*}) ,  &   k= \theta N +1,\cdots,N \end{cases}
\end{align}
for any $0 \leq \theta \leq 1$.
From (\ref{A1}), it follows that $ \frac{1}{N} \sum_{k=1}^{N}  R_{k} ( P_{z,k} , \rho_{z,k} ) = \frac{\theta}{N} \sum_{k=1}^{N}  R_{k} ( P_{x,k}^{*}, \rho_{x,k}^{*} )  +  \frac{1 - \theta}{N} \sum_{k=1}^{N}  R_{k} ( P_{y,k}^{*}, \rho_{y,k}^{*} ) \rightarrow  \theta \mathbb{E} \left[ R_{\nu} ( P_{x,\nu}^{*}, \rho_{x, \nu}^{*}) \right] + (1 - \theta) \mathbb{E} \left[ R_{\nu} ( P_{y,\nu}^{*}, \rho_{y, \nu}^{*}) \right] $ as $N \rightarrow \infty$. 
Also, $ \frac{1}{N} \sum_{k=1}^{N} P_{z, k} \rightarrow  \theta \mathbb{E} [ P_{x, \nu}^{*} ] + (1-\theta) \mathbb{E} [ P_{y, \nu}^{*} ] \leq  \theta \overline{P}_{x}  + (1 - \theta) \overline{P}_{y} $ and $  \frac{1}{N} \sum_{k=1}^{N}  Q_{k}^{\texttt{NL}} ( P_{z,k} , \rho_{z,k} ) \rightarrow  \theta \mathbb{E} [ Q_{\nu}^{\texttt{NL}} (  P_{x, \nu}^{*} , \rho_{x, \nu}^{*} ) ] + (1 - \theta) \mathbb{E} [ Q_{\nu}^{\texttt{NL}} ( P_{y, \nu}^{*} , \rho_{y, \nu}^{*} ) ] \geq \theta Q_{x} + (1-\theta) Q_{y} $ for any $( Q_{x} , \overline{P}_{x} )$ and $( Q_{y} , \overline{P}_{y} )$.
Thus, (P2) satisfies the time-sharing condition.

\subsection{Proof of Lemma \ref{lem_1}}

Following the above procedures with $P_{x,k}^{*} = P_{y,k}^{*} = P_{z,k} = P$, $\forall k$,
it can be shown that (P1) satisfies the time-sharing condition.


\section{Proof of Theorem \ref{thm_1}}
\label{proof_thm_1}


The Lagrange dual function for (P1) is given by \cite{Boyd}
\begin{align}
\label{B1}
& \mathcal{D} ( \lambda )  \nonumber\\
& =  \underset{  0 \leq \rho_{\nu} \leq 1 , \forall \nu  }{\texttt{max}} & ~ \Big\{ \mathbb{E} \left[  R_{\nu} (  P , \rho_{\nu}  )  \right]  + \lambda  \left( \mathbb{E} \left[  Q_{\nu}^{\texttt{NL}} (  P , \rho_{\nu}  )  \right] - Q \right)  \Big\}
\end{align}
where $\lambda$ is the dual variable associated with the average harvested energy constraint of (\ref{P1_const_1}).
To obtain the solution to (P1), in the following, we first determine the optimal $\rho_{\nu}$ by solving (\ref{B1}) given the dual variable $\lambda$.
Then we determine the optimal $\lambda$.

\subsection{Optimal Power Splitting Ratio Given the Dual Variable}

It follows that $  \frac{1}{N} \sum_{k=1}^{N}  R_{k} ( P , \rho_{k} ) \rightarrow \mathbb{E} [  R_{\nu} (  P , \rho_{\nu } )  ] $ and 
$  \frac{1}{N} \sum_{k=1}^{N}  Q_{k}^{\texttt{NL}} ( P , \rho_{k} ) \rightarrow \mathbb{E} [  Q_{\nu}^{\texttt{NL}} (  P , \rho_{\nu } )  ] $ as $N \rightarrow \infty$.
Also, the variables $\{  \rho_{\nu} \}$ are independent since $\{  h_{\nu} \}$ are independent.
Thus, given $\lambda$, the problem of (\ref{B1}) can be decomposed into several (essentially, infinitely many) subproblems, each for one particular fading state (or one summation), as follows \cite{Kang_2}--\cite{Liu_dps}, \cite{Yu}:
\begin{align}
\label{B2}
\underset{ 0 \leq \rho_{\nu}  \leq 1 }{\max}  \quad  \mathcal{L}_{\nu} ( \rho_{\nu} ) , \quad \forall \nu,
\end{align}
where $ \mathcal{L}_{\nu} ( \rho_{\nu} )  =  R_{\nu} ( P, \rho_{\nu} ) + \lambda  Q_{\nu}^{\texttt{NL}} ( P, \rho_{\nu} )  $.
Because the optimization of (\ref{B2}) is nonconvex, the solution needs to be found case by case.
To this end, we exploit the Karush-Kuhn-Tucker (KKT) optimality conditions for (\ref{B2}), which are given by 
$ \frac{ \partial \mathcal{L}_{\nu} } { \partial  \rho_{\nu}  } + \phi_{\nu} -  \psi_{\nu} =  0 $ and $\phi_{\nu} \rho_{\nu}  =  \psi_{\nu} ( 1  -  \rho_{\nu}   ) = 0$ \cite{Boyd},
where $ \phi_{\nu} \geq 0 $ and $\psi_{\nu} \geq 0$ denote the Lagrange multipliers associated with the constraints $\rho_{\nu} \geq 0$ and $\rho_{\nu}  \leq 1 $, respectively.
If $ \rho_{\nu}  = 0 $ (or $ \rho_{\nu} = 1 $), then $\psi_{\nu} = 0$ (or $  \phi_{\nu} = 0 $) from the second condition, and thus, $\frac{ \partial \mathcal{L}_{\nu} } { \partial  \rho_{\nu}  } \leq 0$ (or $\frac{ \partial \mathcal{L}_{\nu} } { \partial  \rho_{\nu}  } \geq 0$) from the first condition.
Similarly, if $ 0 < \rho_{\nu}  < 1  $, then $  \phi_{\nu}= \psi_{\nu} = 0$, and thus, $\frac{ \partial \mathcal{L}_{\nu} } { \partial  \rho_{\nu}  } =  0 $.
Consequently, the KKT conditions are given by
\begin{align}
\label{B4}
\frac{\partial \mathcal{L}_{\nu}}{\partial \rho_{\nu}} & =  \begin{cases}  \leq 0,  &  {\rm if} ~ \rho_{\nu}^{*} = 0 \\ \geq 0,  &  {\rm if} ~ \rho_{\nu}^{*} = 1 \\   = 0,  &  {\rm if} ~ 0 < \rho_{\nu}^{*} < 1  \end{cases}
\end{align}
where $ \rho_{\nu}^{*} $ denotes the solution to (\ref{B2}).
Also, $ \frac{\partial \mathcal{L}_{\nu}}{\partial \rho_{\nu}}  = \lambda \frac{ P_{s} T a h_{\nu} P   }{ 1 - \Omega } \Psi_{\nu} ( P, \rho_{\nu}  ) (  1 - \Psi_{\nu} ( P, \rho_{\nu}  ) )
- \frac{ h_{\nu} P  }{ \left(  (1 - \rho_{\nu} ) h_{\nu} P + \sigma^{2} \right) } $.

First, we consider the case when $\gamma ( h_{\nu} )  \geq \frac{\lambda}{4}$ to derive the solution in the Case 1.
The function $ \Psi_{\nu} ( P, \rho_{\nu}  ) (  1 - \Psi_{\nu} ( P, \rho_{\nu}  ) ) $ has its peak value of $\frac{1}{4}$ at the inflection point $\rho_{\nu} = \min \left\{ \frac{b}{h_{\nu} P} , 1 \right\}$. Thus, it follows that if $ \left. \frac{\partial \mathcal{L}_{\nu}}{\partial \rho_{\nu}} \right|_{ \rho_{\nu} = \min \left\{ \frac{b}{h_{\nu} P} , 1 \right\}} \leq 0 $, or equivalently, $\gamma ( h_{\nu} )  \geq \frac{\lambda}{4}$,
then $ \frac{\partial \mathcal{L}_{\nu}}{\partial \rho_{\nu}} \leq 0 $ for all $ 0 \leq \rho_{\nu} \leq 1$.
Thus, from (\ref{B4}), we have $\rho_{\nu}^{*} = 0$ when $\gamma ( h_{\nu} )  \geq \frac{\lambda}{4}$,
which corresponds to the result of (\ref{rho_NL}) for the Case 1.

Next, we consider the case when $\gamma ( h_{\nu} )  < \frac{\lambda^{\texttt{NL}}}{4}$ to derive the solutions in the Cases 2--5.
From (\ref{B4}), we have
\begin{align}
\label{B6}
\lambda f ( 0 ) \leq g ( h_{\nu} ),  & \quad   {\rm if} ~ \rho_{\nu}^{*} = 0  \\
\label{B7}
\lambda f ( h_{\nu} ) \geq  g ( 0 ),  & \quad   {\rm if} ~ \rho_{\nu}^{*} = 1 .
\end{align}
Also, the value of $0 < \rho_{\nu}^{*} < 1$ is given by the root of the equation $ \frac{\partial \mathcal{L}_{\nu}}{\partial \rho_{\nu}} = 0 $, or equivalently, the equation of (\ref{rho_opt}).
In general, this equation has two roots: one is the local minimum located in the range $\Big[0, \frac{b}{h_{\nu} P} \Big) $ and
the other is the local maximum located in the range $\Big( \frac{b}{h_{\nu} P},  \infty \Big)$. Let $\rho_{{\rm o},\nu}$ denote the local maximum. The solution satisfying $ \frac{\partial \mathcal{L}_{\nu}}{\partial \rho_{\nu}} = 0 $ is given by $\rho_{\nu}^{*} = \rho_{{\rm o},\nu}$.
It follows from (\ref{B4}) that if $ \left. \frac{\partial \mathcal{L}_{\nu}}{\partial \rho_{\nu}} \right|_{\rho_{\nu} = 1} < 0 $, or equivalently, $\lambda f ( h_{\nu} ) <  g ( 0 )$, then the solution $ \rho_{\nu}^{*} $ can be found by solving the equation of (\ref{rho_opt}) over $\left(  \frac{b}{h_{\nu} P} , 1 \right)$. Thus, we have
\begin{align}
\label{B8}
\lambda f ( h_{\nu} ) <  g ( 0 ),  & \quad   {\rm if} ~ 0 < \rho_{\nu}^{*} < 1 .
\end{align}

As can be seen from (\ref{B6})--(\ref{B8}), when $\gamma ( h_{\nu} )  < \frac{\lambda^{\texttt{NL}}}{4}$,
the optimal conditions in (\ref{B4}) depend on the channel power gain $h_{\nu}$ through the functions $f(\cdot)$ and $g(\cdot)$.
In the following, we derive the solution to (\ref{B2}) for the following four possible cases.
\begin{enumerate} [label=\textit{\roman*})]
\item $ \lambda f ( 0 ) \leq g ( h_{\nu} ) $ and $\lambda f ( h_{\nu} )  <  g ( 0 )$: In this case, both the conditions (\ref{B6}) and (\ref{B8}) are satisfied. Thus, the solution can be either $\rho_{\nu}^{*} = 0$ or $\rho_{\nu}^{*} = \rho_{{\rm o}, \nu}$. To obtain the solution, we need to compare the two objective values: $\mathcal{L}_{\nu} ( 0 )$ and $\mathcal{L}_{\nu} ( \rho_{{\rm o}, \nu} )$. If $\mathcal{L}_{\nu} (  \rho_{{\rm o}, \nu} ) > \mathcal{L}_{\nu} ( 0 )$, or equivalently, $R_{\nu} ( P, \rho_{{\rm o}, \nu} )  + \lambda Q_{\nu}^{\texttt{NL}} ( P, \rho_{{\rm o}, \nu} ) >  R_{\nu} ( P , 0 ) $, we have $\rho_{\nu}^{*} =  \rho_{{\rm o}, \nu}$. Otherwise, we have $\rho_{\nu}^{*} = 0$.

\item $ \lambda f ( 0 ) \leq g ( h_{\nu} ) $ and $\lambda f ( h_{\nu} )  \geq  g ( 0 )$: In this case, both the conditions (\ref{B6}) and (\ref{B7}) are satisfied. Thus, the solution can be either $\rho_{\nu}^{*} = 0$ or $\rho_{\nu}^{*} = 1$. To obtain the solution, we need to compare the two objective values: $\mathcal{L}_{\nu} (  0 )$ and $\mathcal{L}_{\nu} (  1 )$. If $\mathcal{L}_{\nu} (  1 ) > \mathcal{L}_{\nu} (  0 ) $, or equivalently, $\lambda Q_{\nu}^{\texttt{NL}} ( P , 1 ) >  R_{\nu} ( P , 0 )$, we have $\rho_{\nu}^{*} = 1$. Otherwise, we have $\rho_{\nu}^{*} = 0$.

\item $ \lambda f ( 0 ) > g ( h_{\nu} ) $ and $\lambda f ( h_{\nu} )  <  g ( 0 )$: In this case, only the condition (\ref{B8}) is satisfied. Thus, we have $\rho_{\nu}^{*} =  \rho_{{\rm o}, \nu}$.

\item $ \lambda f ( 0 ) > g ( h_{\nu} ) $ and $\lambda f ( h_{\nu} )  \geq  g ( 0 )$: In this case, only the condition (\ref{B7}) is satisfied. Thus, we have $\rho_{\nu}^{*} = 1$.
\end{enumerate}

The solutions obtained for the above four cases \textit{i})--\textit{iv}) correspond to the result of (\ref{rho_NL}) for the Cases 2--5, respectively.

\subsection{Optimal Dual Variable}

The optimal value of $\lambda$ can be found by solving the dual problem: $\underset{   \lambda \geq 0  }{\texttt{min}} ~  \mathcal{D} ( \lambda ) $.
From the KKT condition, the optimal $\lambda$ can be determined to satisfy the following completeness slackness condition \cite{Boyd}:
$\lambda \left(   \mathbb{E} \left[  Q_{\nu}^{\texttt{NL}} (  P , \rho_{\nu}^{*}  )  \right] - Q \right)   = 0$ .
If $\lambda = 0$, then $\rho_{\nu}^{*} = 0$ for $\forall \nu$, which is infeasible for (P1).
It thus must be $\lambda > 0$, meaning that the average harvested energy constraint of (\ref{P1_const_1}) must be satisfied with equality.
Therefore, the optimal $\lambda$ is chosen to satisfy $ \mathbb{E} \left[  Q_{\nu}^{\texttt{NL}} (  P , \rho_{\nu}^{*}  )  \right] = Q$.
Let $\lambda^{\texttt{NL}}$ denote the optimal dual variable.
Then, substituting $\lambda^{\texttt{NL}}$ into $\rho_{\nu}^{*}$, and changing the terms $\rho_{\nu}^{*} $ and $\rho_{{\rm o},\nu} $ to $\rho_{\nu}^{\texttt{NL}}$ and $\rho_{{\rm o},\nu}^{\texttt{NL}} $, respectively,
the solution to (P1) can be expressed as in Theorem \ref{thm_1}.

\section{Proof of Theorem \ref{thm_3}}
\label{proof_thm_3}

In this proof, we first transform (P2) into a more tractable form.
Then we derive the optimal solution by solving the transformed problem.

\subsection{Problem Transformation}

Let us define the following new variables: $ \rho_{\nu} P_{\nu} = P_{{\rm EH}, \nu} $ and $ (1 - \rho_{\nu}) P_{\nu} = P_{{\rm ID}, \nu} $, $\forall \nu$,
which denote the power allocations for EH and ID, respectively.
Then the problem (P2) can be equivalently written in the following form:
\begin{subequations}
\label{P3}
\begin{align}
\label{P3_obj}
{\rm (P3):}  \quad  \underset{ \substack{ 0 \leq P_{{\rm ID}, \nu} + P_{{\rm EH}, \nu} \leq P_{\max},  \forall \nu, \\  P_{{\rm ID}, \nu} \geq 0, P_{{\rm EH}, \nu} \geq 0 , \forall \nu  } }{\texttt{max}} & \quad \mathbb{E} \left[  R_{\nu} ( P_{{\rm ID}, \nu} , 0  )  \right]   \\
\label{P3_const_1}
\texttt{s.t.}  &  \quad  \mathbb{E} \left[  Q_{\nu}^{\texttt{NL}} (  P_{{\rm EH}, \nu} , 1  )  \right] \geq Q,  \\
\label{P3_const_2}
&  \quad  \mathbb{E} \left[  P_{{\rm ID}, \nu} + P_{{\rm EH}, \nu}  \right]  \leq P_{\rm avg}.
\end{align}
\end{subequations}
Let $ ( P_{{\rm ID}, \nu}^{\texttt{NL}}, P_{{\rm EH}, \nu}^{\texttt{NL}} ) $ denote the optimal solution to the above problem (P3).
Then, from $ ( P_{{\rm ID}, \nu}^{\texttt{NL}}, P_{{\rm EH}, \nu}^{\texttt{NL}} ) $, the optimal solution $ ( P_{\nu}^{\texttt{NL}}, \rho_{\nu}^{\texttt{NL}} ) $ to (P2) can be obtained by (\ref{P2_sol}).
Thus, in the following, we focus on deriving the solution to (P3).

\subsection{Optimal Power Allocation for ID and EH}

The Lagrange dual function for (P3) is given by
\begin{align}
\label{C1}
& \mathcal{D} ( \lambda , \mu )  =  \underset{ \substack{ 0 \leq P_{{\rm ID}, \nu} + P_{{\rm EH}, \nu} \leq P_{\max},  \forall \nu, \\  P_{{\rm ID}, \nu} \geq 0, P_{{\rm EH}, \nu} \geq 0 , \forall \nu  } }{\texttt{max}}  ~ \Big\{ \mathbb{E} \left[ R_{\nu} ( P_{{\rm ID}, \nu} , 0  ) \right]  + \lambda  \nonumber\\
&  \times \left( \mathbb{E} \left[  Q_{\nu}^{\texttt{NL}} (  P_{{\rm EH}, \nu} , 1  )  \right] - Q \right) + \mu \left(  P_{\rm avg} - \mathbb{E} \left[  P_{{\rm ID}, \nu} + P_{{\rm EH}, \nu}  \right] \right)  \Big \}
\end{align}
where $\lambda$ and $\mu$ denote the dual variables associated with the constraints of (\ref{P3_const_1}) and (\ref{P3_const_2}), respectively.
Similarly as in the case of (P1), it can be shown that the optimal dual variables $\lambda^{\texttt{NL}}$ and $\mu^{\texttt{NL}}$ must be positive, i.e., $\lambda^{\texttt{NL}} > 0$ and $\mu^{\texttt{NL}} > 0$,
meaning that both the constraints of (\ref{P2_const_1}) and (\ref{P2_const_2}) with equality, i.e., $ \mathbb{E} \left[  Q_{\nu}^{\texttt{NL}} (  P_{{\rm EH}, \nu}^{*} , 1  )  \right] = Q$ and $ \mathbb{E} \left[   P_{{\rm ID}, \nu}^{*} + P_{{\rm EH}, \nu}^{*}  \right] =   P_{\rm avg}$, respectively,
where $( P_{{\rm ID}, \nu}^{*}, P_{{\rm EH}, \nu}^{*} )$ denotes the solution to (\ref{C1}).
Thus, in the following, we focus on deriving the solution $( P_{{\rm ID}, \nu}^{*}, P_{{\rm EH}, \nu}^{*}  )$ to (\ref{C1}) given the dual variables $\lambda$ and $\mu$.

We have $  \frac{1}{N} \sum_{k=1}^{N}  R_{k} ( P_{{\rm ID}, k} , 0 ) \rightarrow \mathbb{E} [  R_{\nu} (  P_{{\rm ID}, \nu} , 0 )  ] $,
$  \frac{1}{N} \sum_{k=1}^{N}  Q_{k}^{\texttt{NL}} ( P_{{\rm EH}, k} , 1 ) \rightarrow \mathbb{E} [  Q_{\nu}^{\texttt{NL}} (  P_{{\rm EH}, \nu} , 1 )  ] $,
and $  \frac{1}{N} \sum_{k=1}^{N} (P_{{\rm ID}, \nu} + P_{{\rm EH}, \nu}) \rightarrow \mathbb{E} [  P_{{\rm ID}, \nu} + P_{{\rm EH}, \nu}  ] $ as $N \rightarrow \infty$,
where $P_{{\rm ID}, k} = (1-\rho_{k}) P_{k}$ and $P_{{\rm EH}, k} = \rho_{k} P_{k}$.
Since the variables $\{ P_{{\rm ID}, \nu}, P_{{\rm EH}, \nu} \}$ are independent, given $\lambda$ and $\mu$, the problem of (\ref{C1}) can be decoupled into several (essentially, infinitely many) subproblems, each for one particular fading state (or one summation), as follows \cite{Kang_2}--\cite{Liu_dps}, \cite{Yu}:
\begin{align}
\label{C2}
\underset{ \substack{ 0 \leq P_{{\rm ID}, \nu} + P_{{\rm EH}, \nu} \leq P_{\max}, \\  P_{{\rm ID}, \nu} \geq 0, P_{{\rm EH}, \nu} \geq 0  } }{\max}  \quad  \mathcal{L}_{\nu} ( P_{{\rm ID}, \nu}, P_{{\rm EH}, \nu} ) , \quad  \forall \nu,
\end{align}
where
$ \mathcal{L}_{\nu} ( P_{{\rm ID}, \nu}, P_{{\rm EH}, \nu} )  =  R_{\nu} ( P_{{\rm ID}, \nu} , 0  ) - \mu P_{{\rm ID}, \nu}  + \lambda  Q_{\nu}^{\texttt{NL}} (  P_{{\rm EH}, \nu} , 1  ) - \mu P_{{\rm EH}, \nu} $.
The above problem (\ref{C2}) is solved as follows: first, the optimal $P_{{\rm ID}, \nu}$ is determined (as a function of $P_{{\rm EH}, \nu}$),
and then, the optimal $P_{{\rm EH}, \nu}$ is determined.
Specifically, in (\ref{C2}), the solution $P_{{\rm ID}, \nu}^{*}$ is given by the well-known water-filling power allocation: $P_{{\rm ID}, \nu}^{*} = \min \left\{ \max \left\{ \frac{1}{\mu} - \frac{\sigma^{2}}{h_{\nu}}, 0 \right\} , P_{\max} - P_{{\rm EH}, \nu} \right\}$ \cite[Ch. 5.3.3]{Tse}. Substituting this into (\ref{C2}), the optimization of (\ref{C2}) reduces to the optimization only over the variable $P_{{\rm EH}, \nu}$, from which the solution $P_{{\rm EH}, \nu}^{*}$ can be obtained. The details are given below.

\subsubsection{Optimal Power Allocation for EH When $h_{\nu} \leq  \mu \sigma^{2}$}

When $h_{\nu} \leq  \mu \sigma^{2}$, we have $ P_{{\rm ID}, \nu}^{*} = 0 $. Thus, the optimization of (\ref{C2}) becomes:
\begin{align}
\label{C4}
\underset{ 0 \leq  P_{{\rm EH}, \nu} \leq P_{\max}}{\max}  \quad  \lambda  Q_{\nu}^{\texttt{NL}} (  P_{{\rm EH}, \nu} , 1  ) - \mu P_{{\rm EH}, \nu}.
\end{align}
The solution to the above problem (\ref{C4}) can be obtained by following the similar procedures in \cite[Appendix B]{Kang_3} (the detailed proof is omitted due to the space limit).
Consequently, it is given by $ P_{{\rm EH}, \nu}^{*} =  \left[ P_{{\rm EH}, \nu}^{A} \right]_{0}^{ P_{\max} }$,
where $ \left[ P_{{\rm EH}, \nu}^{A} \right]_{P_{\rm low}}^{ P_{\rm up} } $ is given by (\ref{p_EH_NL_A}).

\subsubsection{Optimal Power Allocation for EH When $h_{\nu} >  \mu \sigma^{2}$}
\label{proof_thm_3_}

When $h_{\nu} >  \mu \sigma^{2}$, it follows that $P_{{\rm ID}, \nu}^{*} = \min \left\{ \frac{1}{\mu} - \frac{\sigma^{2}}{h_{\nu}} , P_{\max} - P_{{\rm EH}, \nu} \right\}$.
Thus, we have the following two sub-cases:
\begin{enumerate} [label=\textit{2-\roman*})]
\item $ 0 \leq P_{{\rm EH}, \nu} \leq P_{{\rm th}, \nu} $: In this sub-case, we have $\frac{1}{\mu} - \frac{\sigma^{2}}{h_{\nu}} \leq P_{\max} - P_{{\rm EH}, \nu}$, and thus, $ P_{{\rm ID}, \nu}^{*} = \frac{1}{\mu} - \frac{\sigma^{2}}{h_{\nu}} $. Substituting this into (\ref{C2}) and dropping the constant term, the optimization of (\ref{C2}) becomes:
    \begin{align}
    \label{C5}
    \underset{ 0 \leq  P_{{\rm EH}, \nu} \leq P_{{\rm th}, \nu} }{\max}  \quad  \lambda  Q_{\nu}^{\texttt{NL}} (  P_{{\rm EH}, \nu} , 1  ) - \mu P_{{\rm EH}, \nu}.
    \end{align}
    Similarly as in (\ref{C4}), the solution to (\ref{C5}) is given by $ P_{{\rm EH}, \nu}^{*} =  \left[ P_{{\rm EH}, \nu}^{A} \right]_{0}^{ P_{{\rm th}, \nu} }$.

\item $ P_{{\rm th}, \nu} < P_{{\rm EH}, \nu} \leq P_{{\rm th}, \nu} $: In this sub-case, we have $\frac{1}{\mu} - \frac{\sigma^{2}}{h_{\nu}} \leq P_{\max} - P_{{\rm EH}, \nu}$, and thus, $ P_{{\rm ID}, \nu}^{*} = P_{\max} - P_{{\rm EH}, \nu}$. Substituting this into (\ref{C2}) and dropping the constant term, the optimization of (\ref{C2}) becomes:
    \begin{align}
    \label{C6}
    \underset{ 0 \leq  P_{{\rm EH}, \nu} \leq P_{{\rm th}, \nu} }{\max}  \quad  R_{\nu} ( P_{\max} - P_{{\rm EH}, \nu} , 0  ) +  \lambda  Q_{\nu}^{\texttt{NL}} (  P_{{\rm EH}, \nu} , 1  ).
    \end{align}
    Following the similar procedures in Appendix \ref{proof_thm_1}, it can be shown that the solution to (\ref{C6}) is given by $ P_{{\rm EH}, \nu}^{*} =  \left[ P_{{\rm EH}, \nu}^{B} \right]_{ P_{{\rm th}, \nu} }^{ P_{\max} }$,
    where $ \left[ P_{{\rm EH}, \nu}^{B} \right]_{P_{\rm low}}^{ P_{\rm up} } $ is given by (\ref{p_EH_NL_B}).
\end{enumerate}

To obtain the solution when $h_{\nu} >  \mu \sigma^{2}$, we need to compare the two objective values of (\ref{C2}) for the above two sub-cases:
one is given by $ \mathcal{L}_{\nu} \left( \frac{1}{\mu} - \frac{\sigma^{2}}{h_{\nu}}, \left[ P_{{\rm EH}, \nu}^{A} \right]_{0}^{ P_{{\rm th}, \nu} } \right) =  \log_{2} \left( \frac{h_{\nu}}{\lambda \sigma^{2}} \right) + \frac{\lambda \sigma^{2}}{h_{\nu}} - 1 + \lambda  Q_{\nu}^{\texttt{NL}} (  P_{{\rm EH}, \nu} , 1  ) - \mu \cdot \left[ P_{{\rm EH}, \nu}^{A} \right]_{0}^{ P_{{\rm th}, \nu} }$ and the other is given by $ \mathcal{L}_{\nu} \left( P_{\max} - \left[ P_{{\rm EH}, \nu}^{B} \right]_{ P_{{\rm th}, \nu} }^{ P_{\max} } , \left[ P_{{\rm EH}, \nu}^{B} \right]_{ P_{{\rm th}, \nu} }^{ P_{\max} } \right) =  R_{\nu} ( P_{\max} - P_{{\rm EH}, \nu} , 0  ) +  \lambda  Q_{\nu}^{\texttt{NL}} (  P_{{\rm EH}, \nu} , 1  ) - \mu P_{\max}$.
If the former is larger than the latter, we have $( P_{{\rm ID}, \nu}^{*}, P_{{\rm EH}, \nu}^{*} ) = \left( \frac{1}{\mu} - \frac{\sigma^{2}}{h_{\nu}}, \left[ P_{{\rm EH}, \nu}^{A} \right]_{0}^{ P_{{\rm th}, \nu} } \right)$.
Otherwise, we have $( P_{{\rm ID}, \nu}^{*}, P_{{\rm EH}, \nu}^{*} ) = \Big( P_{\max} - \left[ P_{{\rm EH}, \nu}^{B} \right]_{ P_{{\rm th}, \nu} }^{ P_{\max} } , \allowbreak \left[ P_{{\rm EH}, \nu}^{B} \right]_{ P_{{\rm th}, \nu} }^{ P_{\max} } \Big)$.




\ifCLASSOPTIONcaptionsoff
  \newpage
\fi



\end{document}